\documentclass[twocolumn,trackchanges]{aastex631}

\usepackage{CJK}

\newcommand{\petit}{\texttt{petitRADTRANS}}

\newcommand{\dynesty}{\texttt{dynesty}}
\newcommand{\teff}{T$_{\rm eff}$}

\newcommand{\kms}{$\rm km\ s^{-1}$ }
\newcommand{\caltech}{Department of Astronomy, California Institute of Technology, Pasadena, CA 91125, USA}
\newcommand{\gps}{Division of Geological \& Planetary Sciences, California Institute of Technology, Pasadena, CA 91125, USA}
\newcommand{\ucsc}{Department of Astronomy \& Astrophysics, University of California, Santa Cruz, CA95064, USA}
\newcommand{\keck}{W. M. Keck Observatory, 65-1120 Mamalahoa Hwy, Kamuela, HI, USA}
\newcommand{\ucla}{Department of Physics \& Astronomy, 430 Portola Plaza, University of California, Los Angeles, CA 90095, USA}
\newcommand{\jpl}{Jet Propulsion Laboratory, California Institute of Technology, 4800 Oak Grove Dr.,Pasadena, CA 91109, USA}
\newcommand{\ucsd}{Center for Astrophysics and Space Sciences, University of California, San Diego, La Jolla, CA 92093}

\received{\today}
\revised{}
\accepted{}

\shorttitle{Thermal Emission of WASP-33b}
\shortauthors{Finnerty et al.}
\graphicspath{{./}{}}

\begin{document}
\begin{CJK*}{UTF8}{gbsn}

\title{Keck/KPIC Emission Spectroscopy of WASP-33b}

\correspondingauthor{Luke Finnerty}
\email{lfinnerty@astro.ucla.edu}

\author[0000-0002-1392-0768]{Luke Finnerty}
\affiliation{\ucla}

\author{Tobias Schofield}
\affiliation{\caltech}

\author[0000-0003-1399-3593]{Ben Sappey}
\affiliation{\ucsd}

\author[0000-0002-6618-1137]{Jerry W. Xuan}
\affiliation{\caltech}

\author[0000-0003-2233-4821]{Jean-Baptiste Ruffio}
\affiliation{\caltech}

\author[0000-0003-0774-6502]{Jason J. Wang (王劲飞)}
\altaffiliation{51 Pegasi b Fellow}
\affiliation{\caltech}

\author[0000-0001-8953-1008]{Jacques-Robert Delorme}
\affiliation{\keck}
\affiliation{\caltech}

\author{Geoffrey A. Blake}
\affiliation{\gps}

\author{Cam Buzard}
\affiliation{Division of Chemistry and Chemical Engineering, California Institute of Technology, Pasadena, CA 91125, USA}

\author[0000-0002-0176-8973]{Michael P. Fitzgerald}
\affiliation{\ucla}

\author{Ashley Baker}
\affiliation{\caltech}

\author{Randall Bartos}
\affiliation{\jpl}

\author{Charlotte Z. Bond}
\affiliation{UK Astronomy Technology Centre, Royal Observatory, Edinburgh EH9 3HJ, United Kingdom}

\author[0000-0003-4737-5486]{Benjamin Calvin}
\affiliation{\caltech}
\affiliation{\ucla}

\author{Sylvain Cetre}
\affiliation{\keck}

\author{Greg Doppmann}
\affiliation{\keck}

\author{Daniel Echeverri}
\affiliation{\caltech}

\author[0000-0001-5213-6207]{Nemanja Jovanovic}
\affiliation{\caltech}

\author[0000-0002-4934-3042]{Joshua Liberman}
\affiliation{\caltech}

\author[0000-0002-2019-4995]{Ronald A. L\'opez}
\affiliation{\ucla}

\author[0000-0002-0618-5128]{Emily C. Martin}
\affiliation{\ucsc}

\author{Dimitri Mawet}
\affiliation{\caltech}
\affiliation{\jpl}

\author{Evan Morris}
\affiliation{\ucsc}

\author{Jacklyn Pezzato}
\affiliation{\caltech}

\author[0000-0001-5610-5328]{Caprice L. Phillips}
\affiliation{Department of Astronomy, The Ohio State University, 100 W 18th Ave, Columbus, OH 43210 USA}


\author{Sam Ragland}
\affiliation{\keck}

\author{Andrew Skemer}
\affiliation{\ucsc}

\author{Taylor Venenciano}
\affiliation{Physics and Astronomy Department, Pomona College, 333 N. College Way, Claremont, CA 91711, USA}

\author{J. Kent Wallace}
\affiliation{\jpl}

\author[0000-0003-0354-0187]{Nicole L. Wallack}
\affiliation{\gps}

\author[0000-0002-4361-8885]{Ji Wang (王吉)}
\affiliation{Department of Astronomy, The Ohio State University, 100 W 18th Ave, Columbus, OH 43210 USA}

\author{Peter Wizinowich}
\affiliation{\keck}



\begin{abstract}

We present Keck/KPIC high-resolution ($R\sim35,000$) $K$-band thermal emission spectroscopy of the ultra-hot Jupiter WASP-33b. The use of KPIC's single-mode fibers greatly improves both blaze and line-spread stabilities relative to slit spectrographs, enhancing the cross-correlation detection strength. We retrieve the dayside emission spectrum with a nested sampling pipeline which fits for orbital parameters, the atmospheric pressure-temperature profile, and molecular abundances.We strongly detect the thermally-inverted dayside and measure mass-mixing ratios for CO ($\log\rm CO_{MMR} = -1.1^{+0.4}_{-0.6}$), H$_2$O ($\log\rm H_2O_{MMR} = -4.1^{+0.7}_{-0.9}$) and OH ($\log\rm OH_{MMR} = -2.1^{+0.5}_{-1.1}$), suggesting near-complete dayside photodissociation of H$_2$O. The retrieved abundances suggest a carbon- and possibly metal-enriched atmosphere, with a gas-phase C/O ratio of $0.8^{+0.1}_{-0.2}$, consistent with the accretion of high-metallicity gas near the CO$_2$ snow line and post-disk migration or with accretion between the soot and H$_2$O snow lines.  We also find tentative evidence for $\rm ^{12}CO/^{13}CO \sim 50$, consistent with values expected in protoplanetary disks, as well as tentative evidence for a metal-enriched atmosphere (2--15$\times$ solar). These observations demonstrate KPIC's ability to characterize close-in planets and the utility of KPIC's improved instrumental stability for cross-correlation techniques.

\end{abstract}



\section{Introduction} \label{sec:intro}

While large surveys have started to provide population-level insight into the physical and orbital properties of exoplanet systems, the details of exoplanet atmospheres, including composition, formation history, and 3D thermal structure, remain uncertain, with known trends suggesting substantial diversity \citep{mansfield21}. Improved knowledge of hot Jupiter atmospheres in particular may be a key piece to understanding the process of giant planet formation. Different formation scenarios are expected to produce differences in both relative atmospheric abundances \citep[e.g.][]{oberg2011, booth2017} and/or overall metallicity \citep[e.g.][]{espinoza2017,madhusudhan2017, cridland2019}. Recent observations \citep{Pelletier2021, Line2021} have retrieved abundances which appear to conflict with the predictions from some of these theories, particularly the expectation of an inverse correlation between metallicity and C/O ratio \citep{espinoza2017, cridland2019}. These abundances may be explained by pebble drift \citep{booth2017}, but these discrepancies suggest our understanding of either the planet formation process or atmospheric retrievals may be substantially incomplete. 
	
Ultra-hot Jupiters (UHJs), with equilibrium temperatures $\rm T_{eq} > 2000\ K$, are a particularly interesting case for understanding planet formation processes.Two of the three hottest-known UHJs, KELT-9b and WASP-33b, orbit A-type stars on significantly misaligned orbits \citep{gaudi2017, collier2010}, suggesting a dynamical history of eccentric Kozai-Lidov effects \citep{naoz2011}, though no additional planets are currently known in either system. Several A/B-type stars (e.g. HR 8799, $\beta$ Pic) have been found to host one or more massive planets on wide orbits, raising the possibility that UHJs are an alternative evolutionary outcome of such systems. UHJ atmospheres may thus hold indications of when and how planet migration occurred.
	
Understanding the formation of UHJs will also require understanding their extreme atmospheric chemistry. Commonly-used chemical equilibrium assumptions are expected to fail in the case of UHJs, particularly on the dayside, where extreme UV fluxes are expected to produce thermal inversions, drive mass loss, dissociate molecules, and even ionize some transition metals \citep{wyttenbach2020, fu2022, yan2021, nugroho2021, casasayas2019}. These effects must be incorporated into 3D models in order to accurately model global energy transport \citep[e.g.][]{roth2021}. Transitions in phase/ionization state are expected on the nightside, when UV fluxes decline, though this will depend on the details of thermal redistribution between the hemispheres. Indications of rainout have been reported from analyses of transit observations of some UHJs \citep{ehrenreich2020, wardenier2021, johnson2023}, but such observations have difficulty in unambiguously localizing a signal to a particular longitudinal range due to the complexity of 3D atmospheric circulation \citep{savel2022}. Even the possibilities of photodissociation and rainout processes suggest current global circulation models (GCMs) coupled to a chemical-equilibrium radiative transfer framework are inadequate for modeling UHJ atmospheres, though understanding the modes of these failures and when they occur will require additional observational constraints \citep{pluriel20222}. Understanding such modeling issues is critical for successfully interpreting chemical abundances as indicators of planetary formation and evolutionary history.
	
Recent advances in both analysis techniques and instrumentation have positioned high-resolution cross-correlation spectroscopy (HRCCS) as a promising method for obtaining constraints on the thermal and chemical properties of hot Jupiter atmospheres. This technique uses the change in the radial velocity of a close-in planet over several hours to isolate velocity-variable planetary spectral features from quasi-fixed stellar features using a cross-correlation (CC) template believed to match the planet and previously known orbital properties. Initially used to detect molecules such as CO and H$_2$O in hot Jupiter atmospheres \citep[e.g.][]{snellen2010, brogi2012, lockwood2014, buzard2020}, \citet{brogi2019} developed a log-likelihood (logL) function that enables Bayesian retrievals of high-resolution observations, using forward models to produce CC templates which better match observations. Using this approach, \citet{Pelletier2021} retrieved a CO abundance and pressure-temperature (PT) profile for the hot Jupiter $\tau$ Boo A b, and \citet{Line2021} were able to retrieve both CO and H$_2$O abundances for WASP-77A b, yielding the first robust C/O ratio measurement for an unresolved planetary companion from high-resolution spectroscopy. Key to these retrievals are fast radiative transfer tools, which can quickly produce high-resolution spectra for planets with arbitrary molecular abundances and PT profiles, obviating assumptions about atmospheric chemistry or bulk composition.
	
Concurrent with these improvements in data analysis, new instruments are being developed and deployed with design features better suited for CC-based techniques than previous facilities. In particular, \citet{Rasmussen2021} identified correlated noise and minor blaze function variations as having major impacts on the final detection strength when using these techniques, which can be improved by using highly stable instruments. The use of single-mode fibers (SMFs) in Keck/KPIC \citep{kpic} accomplishes this for the NIRSPEC \citep{nirspec, nirspecupgrade, nirspecupgrade2} high-resolution ($R\sim35000$) spectrograph on Keck. While the the Keck AO system and fiber coupling losses reduce overall throughput by a factor of $\sim7$ compared with a seeing-limited slit spectrograph in KPIC phase I \citep{kpic}, the SMFs offer an ultra-stable line-spread function, as well as improvements in the blaze function and wavelength solution stabilities (though systematics may persist below the \kms level, well below the NIRSPEC resolution). Our analysis of KPIC phase I data showed no indications of intra-night variations $>1$ \kms in either the blaze function or wavelength solution. While \citet{Finnerty2022} reported no apparent variation at the $\sim1\%$ level in the KPIC LSF over phase I,  KPIC suffers from a time-varying fringing effect which must be corrected (see Section \ref{sec:retrieval}), caused by interference between two pickoff dichroics \citep{Finnerty2022}. KPIC is a prototype instrument with ongoing hardware upgrades, which will correct this in the future. 
	
In this paper, we present Keck/KPIC phase I observations of the UHJ WASP-33b covering the dayside hemisphere. We use a free-retrieval approach to recover orbital velocity parameters, chemical abundances, and pressure-temperature (PT) profiles. Section \ref{ssec:props} presents properties of the system and previous results for atmospheric characterization of WASP-33b, followed by a presentation of our KPIC observations in Section \ref{ssec:obs}. We discuss our retrieval procedure and choice of parameters/priors in Section \ref{sec:retrieval}, including a verification of our retrieval setup using simulated datasets with known inputs. The results of our retrievals for WASP-33b are presented in Section \ref{sec:res}. We discuss these results in the context of previous studies of UHJ atmospheres and planet formation predictions in Section \ref{sec:disc}. Section \ref{sec:conc} summarizes our results.

\section{Observations and Data Reduction} \label{sec:obs}
\subsection{Target Properties}\label{ssec:props}
\begin{deluxetable}{ccc}
    \tablehead{\colhead{Property} & \colhead{Value} & \colhead{Ref.}}
    \startdata
        & \textbf{WASP-33} & \\
        \hline
        RA & 02:26:51.06 & \citet{gaiaedr3} \\
        Dec & +37:33:01.7 & \citet{gaiaedr3} \\
        Spectral Type & A5 & \citet{grenier1999} \\
        $K_{mag}$ & 7.47 & \citet{cutri2003} \\
        Mass & 1.495 $\rm M_\odot$ & \citet{collier2010} \\
        Radius & 1.44 $\rm R_\odot$ & \citet{collier2010} \\
        \teff & 7400 K & \citet{collier2010} \\
        $v\sin i$ & 86.6 \kms & \citet{johnson2015} \\
        $v_{sys}$ & -9.2 \kms & \citet{gontcharov2006} \\
        Age & $<400$ Myr& \citet{collier2010} \\
        $z$ & +0.1 dex & \citet{collier2010} \\
        \smallskip \\
         & \textbf{WASP-33b} & \\
        \hline
        Period & 1.2198697 days & \citet{smith2011} \\
        $\rm t_{transit}$ & JD 2459509.9195 & \citet{smith2011} \\
        $a$ & 0.02558 AU & \citet{smith2011} \\
        $i$ & 84.6$^\circ$ & \citet{stephan2022} \\
        $K_p$ & 226 \kms & Est.  \\
        Mass & 2.8 $\rm M_J$ & \citet{vonessen2014} \\
        Radius & 1.5 $\rm R_J$ & \citet{collier2010} \\
        $\rm T_{eq}$ & 3300 K & \citet{smith2011} \\
        C/O & $0.8^{+0.1}_{-0.2}$ & This work \\
    \enddata
    \caption{Star and planet properties for the WASP-33 system. Transit time was estimated using the NASA exoplanet ephemeris service. We estimate $K_p$ from the semi-major axis and orbital period, which gives a slightly lower value than the $\sim230$ \kms reported in \citet{yan2019} and \citet{nugroho2021}. \citet{vansluijs2022} reported $K_p$ values ranging from 220 to 240 \kms, depending on the observed orbital phase and cross-correlation model. In retrievals, we use $K_p$ priors that encompass the full range of reported values.} 
    \label{tab:props}
\end{deluxetable}

WASP-33 (HD 15082) is a bright \citep[$K$ = 7.47,][]{cutri2003}, rapidly rotating ($v \sin i = 86$ \kms) A5 star \citep{collier2010, johnson2015}. WASP-33b was first discovered by \citet{christian2003} in SuperWASP \citep{pollacco2006} transit photometry with a 1.22-day orbital period. Subsequent radial velocity measurements obtained a mass of 2.8 $\rm M_J$ \citep{vonessen2014}. The estimated planet mass and host star spectral type are similar to the HR 8799 system \citep{gozdziewski2020}, though the WASP-33b semimajor axis is more than 10$^3$ times smaller. 

As a transiting planet around a rapid rotator, WASP-33b is an ideal target for observing the Rossiter-McLaughlin Effect \citep[RME;][]{ohta2005}. Observations by \citet{collier2010} found that WASP-33b has a near-polar retrograde orbit, with subsequent observations measuring orbital precession and the stellar quadrupole moment \citep{johnson2015, stephan2022}. The extremely small semimajor axis and near-polar orbit suggest the possibility of eccentric Kozai-Lidov dynamical effects \citep{naoz2011}, though there are presently no additional planets known in the system to act as perturbers. This system architecture and misaligned orbit is extremely similar to the UHJ KELT-9b \citep{gaudi2017}, which may be a result of a common formation pathway for both systems. 

WASP-33b has been a frequent target for previous atmospheric characterization studies. Near-IR dayside observations have detected CO and OH in emission, but found only weak signs of H$_2$O, indicative of high-temperature thermochemistry \citep{nugroho2021, vansluijs2022, yan2022}. Optical observations have detected a number of oxides and atomic species, including AlO \citep{vonessen2019}, Si I \citep{cont2022}, Fe I \citep{cont2021, herman2022}, Ti I, and V I \citep{cont2022opt}, while transit observations of Balmer lines suggest an extended escaping hydrogen envelope driving mass loss at a rate of $\sim 10^{12}\rm\ g/s$ \citep{yan2021}. These observations have focused on dayside longitudes or the terminators, with phase curve observations suggesting inefficient redistribution of heat to the nightside \citep{zhang2018, herman2022}. These findings paint the picture of WASP-33b as an extreme ultra-hot Jupiter, with mass loss, photochemistry, and dramatic day/night differences.

WASP-33 is a rapid rotator, with rotational broadening and gravity darkening effects significantly impacting the stellar spectrum. WASP-33 is also a $\delta$ Scuti pulsator \citep{herrero2011}. We discuss our treatment of these effects in Section \ref{ssec:templates}.

\subsection{Observations}\label{ssec:obs}
\begin{figure}
    \centering
    \includegraphics[width=1.0\columnwidth]{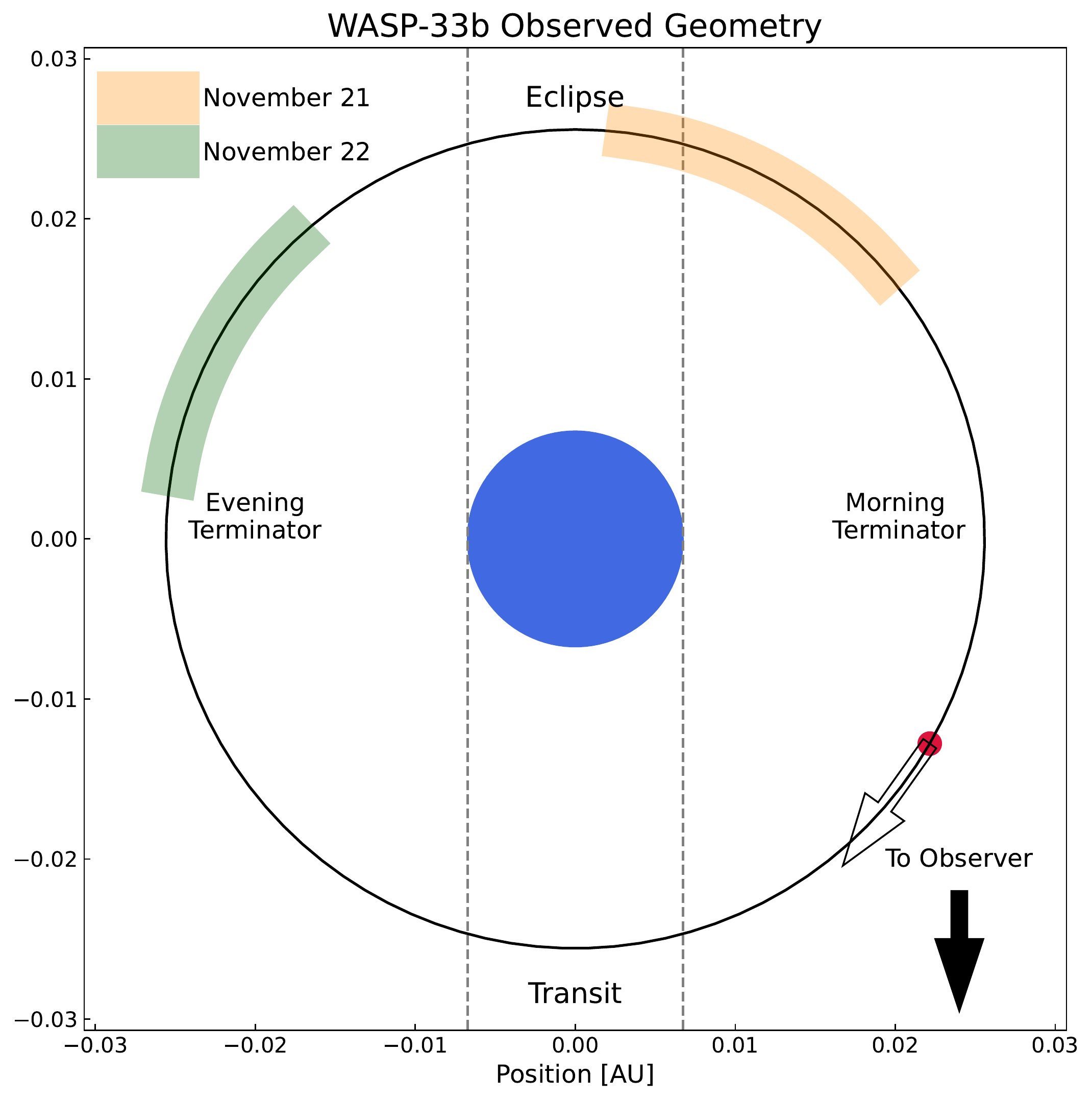}
    \caption{Observed orbital geometry of the WASP-33 system. The diagram is drawn to-scale with the planet orbit face-on and observed orbital phases shaded by observation date. The ellipticity of the star is neglected, and the planet's motion is drawn clockwise to reflect the retrograde orbit. The 2021 November 21 observations probe dayside thermal emission and provide a planet-free stellar spectrum during eclipse. The 2021 November 22 observations targeted the afternoon/dusk region but suffered from poor conditions and are therefore omitted from analysis.}
    \label{fig:obsgeo}
\end{figure}

\begin{deluxetable*}{cccccc}
    \tablehead{\colhead{UT Date} & \colhead{Wall Time [min]} & \colhead{Integration Time [min]} & \colhead{Orbital Phase} & \colhead{Airmass} & \colhead{Throughput} }
    \startdata
        2021 November 21 & 192 & 144 & 0.52-0.64 & 1.08-1.05-1.21 & $\sim2.3\%$ \\
        2021 November 22 & 185 & 130 & 0.28-0.38 & 1.31-1.05 & $<1\%$ \\
    \enddata
    \caption{Summary of presented KPIC observations of WASP-33. The airmass column lists airmass at the start of the observation, the minimum airmass, and the airmass at the end of the observing sequence. For both nights, a significant airmass range was observed, which is beneficial for PCA-based atmospheric detrending.  Instrument performance on the first night was excellent for KPIC phase I, with throughputs consistently above $2\%$ giving per-channel SNR of $\sim100$ in the extracted 1D spectra for each frame. Poor AO correction on the second night led to substantially lower throughput, and we omit these data from our analysis.}
    \label{tab:obs}
\end{deluxetable*}

We observed WASP-33 with Keck/KPIC \citep{kpic} on UT dates 2021 November 21 and 2021 November 22. KPIC (Keck Planet Imager and Characterizer) is a series of upgrades to Keck II/NIRSPEC \citep{nirspec, nirspecupgrade, nirspecupgrade2} and Keck II AO to enable high-resolution ($R\sim35,000$) diffraction-limited spectroscopy for Keck. Primarily intended for spectroscopic follow-up of directly imaged companions, the use of single-mode fibers offers a factor of $\sim$100 reduction in sky background, as well as long-term blaze and line-spread function (LSF) stability. \citet{Rasmussen2021} demonstrated that even minor variations in the LSF or blaze function can significantly impact detection of close-in planets through cross-correlation techniques. Previously, \citet{Finnerty2021} found that cross-correlation techniques are much more severely impacted by minor systematic errors compared with modest increases in Gaussian noise. These findings suggest that the improvements in LSF/blaze stability with KPIC may offset the increase in Poisson noise due to lower throughput in high-resolution cross-correlation applications compared with seeing-limited slit-fed spectrographs. We note that the KPIC phase I throughput was comparable to AO-fed NIRSPEC \citep{Finnerty2022} and that KPIC phase II significantly exceeds the throughput performance of NIRSPAO \citep{echeverri2022}.

Our observations are summarized in Table \ref{tab:obs}. For all observations, we used an ABBA pattern nodding between KPIC science fibers 1 and 2. These fibers had the highest throughputs of the four science fibers in KPIC phase I (tested at the start of each night), and the nodding allows us to effectively subtract sky emission features and thermal emission from warm front-end optics on the Keck AO bench. We obtained 90-s exposures at each nod position, giving our observations a time resolution of approximately four minutes after read time/overheads and coadding frames. We chose a time resolution under five minutes based on simulations showing that the orbital motion of WASP-33b would begin to significantly broaden the final cross-correlation peak for longer timescales (known as Doppler smearing).  Figure \ref{fig:obsgeo} presents a diagram of the system with the observed orbital phases shaded by observation date. 

The 2021 November 21 observations began during secondary eclipse, providing a stellar spectrum with no planet contribution. Observations continued post-eclipse for just under three hours, covering roughly noon to mid-morning longitudes. Conditions were good, with throughput from the top of Earth's atmosphere through detection peaking at consistently around 2.5\%.

We attempted to obtain additional observations on 22 November covering late-evening to afternoon longitudes. However, high clouds caused significant extinction and poor AO performance, leading to top-of-atmosphere throughput consistently below 1\%. We therefore omitted these observations from subsequent cross-correlation and retrieval analysis.

\subsection{Data Reduction}\label{ssec:red}
We began by fitting the echellogram trace location and width for each fiber and order from observations of a bright calibrator star (typically the telluric standard for the wavelength calibrator, discussed below). Because the KPIC fiber-extraction unit (FEU) does not move relative to the spectrometer backend during science operations, this only needs to be done once per night. We then AB subtracted the science frames (taken in an ABBA sequence) to remove sky emission and thermal background. Spectral extraction was performed by summing over slices in the detector-Y direction (which is well aligned with the cross-dispersion direction) centered on the best-fit trace center with a width five times the standard deviation of the best-fit Gaussian. 

Each fiber was individually wavelength-calibrated by comparing to a radial velocity standard following the procedure described in Section 3.2 of \citet{wang2021}. In brief, we observed a late-type giant star (HIP\,95771 in the case of the WASP-33 observations) at the start of each night and fit the observed spectrum with a combined telluric spectrum generated using PSG \citep{psg} and stellar model from the PHOENIX library \citep{phoenix}. Using a late-type star greatly increases the number of lines used for calibration compared with using only the sparse $K$-band telluric absorption features, enabling a more robust calibration.

We next resampled all the extracted spectra from science fiber 1 onto the wavelength grid obtained for science fiber 2 using a cubic spline. While each fiber is stable in detector location and wavelength over a single night, manufacturing and alignment differences can lead to small variations between fibers in both fringing and LSF that we wish to average. We arrange the extracted spectra for each order into a times series, giving $n_\mathrm{order}$ arrays each of shape $n_\mathrm{nods} \times n_\mathrm{channels}$. These arrays, along with the Julian date/time at the midpoint of each nod pair and the corresponding barycentric radial velocity to WASP-33 from Keck, are the inputs to our retrieval code. We omit the first three orders (blueward of 2.1 $\mu\rm m$) due to strong telluric contamination, and the following two orders due to wavelength calibration inaccuracies, leaving four NIRSPEC orders spanning $\sim2.2-2.5\ \mu$m, with significant gaps. Additional post-processing performed as part of the retrieval is discussed in Section \ref{ssec:proc}. 

\section{Atmospheric Retrieval} \label{sec:retrieval}

Our approach to atmospheric retrieval is similar to that of \citet{Line2021}, with the most significant differences being our use of a negative injection of the proposed planet model (see Section \ref{ssec:proc}) prior to the principal component analysis and the use of \petit\ \citep{prt:2019, prt:2020} for radiative-transfer calculations. In brief, we use \petit\ to generate simulated time-series spectra matched to our detrended observations, and calculate a log-likelihood following \citet{brogi2019}. We use this log-likelihood function to find best-fit orbital and atmospheric parameters using nested sampling \citep{skilling2004}, specifically the \dynesty\ \citep{speagle2020} implementation. We discuss the details of this procedure below. 

\subsection{Retrieval Procedure}
\subsubsection{Data Processing}\label{ssec:proc}
\begin{figure*}
    \centering
    \includegraphics[width=2.1\columnwidth]{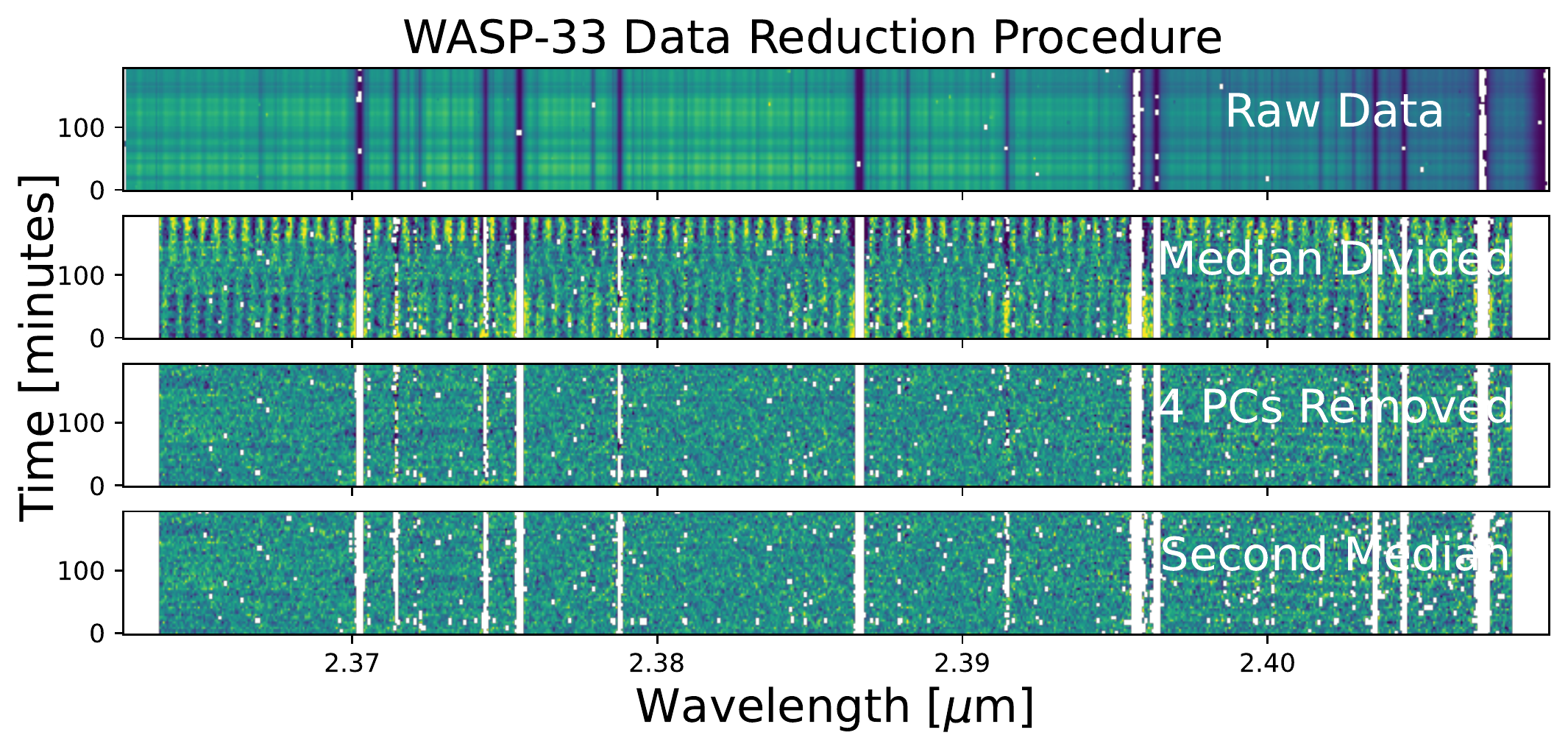}
    \caption{Data processing steps for one order. Top panel shows the raw data time series for one order, with evident tellurics and frame-to-frame variations in total flux. Second panel shows the data after scaling, dividing by a median spectrum, and masking values differing from the median by $>8\%$ (approximately 10 times the median absolute deviation). Telluric and stellar features are effectively removed at this step, but a time-varying fringe pattern is clearly visible. Third panel shows data after removing the first four principal components. Occasionally weak temporally-fixed features persist after PCA. The fourth panel shows the data after a second median division to remove any such features and masking values differing from the median by $>4\%$ (approximately 6 times the median absolute deviation)}
    \label{fig:reduction}
\end{figure*}

At the end of the data reduction process, we can write each extracted spectrum as:
\begin{equation}
    O(\lambda, t) = [S(\lambda) + P(\lambda, t)] T(\lambda, t)  B(\lambda) F(\lambda)  F(\lambda,t)
\end{equation}
Where $O(\lambda, t)$ is the observed flux in each exposure, $S(\lambda)$ is the stellar flux and associated photon noise, including rotational broadening as well as limb and gravity darkening. $P(\lambda, t)$ is the planet flux and associated photon noise, which varies in time both due to the Doppler shift from the planet's orbital motion and due to variations in the visible longitudinal range. $T(\lambda, t)$ describes telluric transmission, which varies in time as the airmass and precipitable water vapor change. $B(\lambda)$ encapsulates the spectrograph blaze function and flat-field effects and is stable over a given night. Finally, we describe the fringing with both a static term $F(\lambda)$, arising from the NIRSPEC entrance window, and a time-varying term $F(\lambda,t)$ which is the result of the KPIC tracking camera dichroic and the pyramid wavefront sensor dichroic producing an etalon with a varying angle of incidence \citep{Finnerty2022}.

Because the projected radial velocity variation of WASP-33b over each observation sequence is substantially more than 3$\times$ the velocity resolution of NIRSPEC, taking a median of the time series should leave only the terms which do not vary significantly over time. However, frame-to-frame variations in coupling efficiency as a result of variable AO correction require that we first divide each spectrum by its median (see Figure \ref{fig:reduction}, top panel) before taking the median over the time series. We can then write the time-series median spectrum $M(\lambda)$ as:
\begin{equation}
    M(\lambda) = \bar{S}(\lambda)\times \bar{T}(\lambda)\times B(\lambda)\times F(\lambda)
\end{equation}
Where $\bar{S}(\lambda)$ is the average stellar spectrum and  $\bar{T}(\lambda)$ represents the time-averaged telluric spectrum. We can then scale all spectra in the time series by the median and divide to obtain:
\begin{equation}\label{eqn:meddiv}
    \frac{O(\lambda,t)}{M(\lambda)} = \left(1+\frac{P(\lambda,t)}{\bar{S}(\lambda)}\right)\frac{T(\lambda,t)}{\bar{T}(\lambda)}\times F(\lambda, t)
\end{equation}
This is shown in the second panel of Figure \ref{fig:reduction}. Both $T(\lambda,t)/\bar{T}(\lambda)$ and $F(\lambda, t)$ vary around unity, so we can combine these terms and rewrite them as:
\begin{equation}
    \frac{T(\lambda,t)}{\bar{T}(\lambda)}\times F(\lambda, t) = 1+\delta f(\lambda, t)
\end{equation}
Using the above expression to rewrite equation \ref{eqn:meddiv} and expanding gives:
\begin{equation}
     \frac{O(\lambda,t)}{M(\lambda)} = 1 + \frac{P(\lambda,t)}{\bar{S}(\lambda)} + \frac{P(\lambda,t)}{\bar{S}(\lambda)}\delta f(\lambda,t) + \delta f(\lambda, t)
\end{equation}

At this point, a number of approaches have been used to remove the time-varying systematic noise/telluric residual term $\delta f(\lambda, t)$ (note that the cross term can generally be taken to be smaller than the others). One approach, used in e.g. \citet{brogi2019, Pelletier2021}, is to fit and divide a low-order polynomial to the time series of each spectral channel. This is effective when the temporal variation is slow, such as airmass-induced variations, and may be followed by additional processing such as PCA. For KPIC data, the time-varying fringe cannot be corrected with this approach, and we therefore do not use it. 

An additional or alternative technique used in e.g., \citet{Pelletier2021, Line2021}, is to perform principal component analysis (PCA) on the time series. PCA is a dimensionality reduction technique which projects the observed spectra onto a basis which maximizes the time-series variation in each successive component. This allows most of the variation in the time series to be encapsulated in the first few principal components, which can then be ``zeroed-out'' via a projection matrix in order to eliminate the variation. While this effectively eliminates most of the time-varying tellurics and fringing, it also removes/distorts some of the planet signal we are attempting to retrieve. This can be minimized by using the smallest number of principal components (PCs) needed to eliminate tellurics/fringing, as the planet signal is generally much lower amplitude and therefore should be more present in higher PCs.

The distortion of the spectrum caused by PCA can then be estimated by using the calculated PCs to apply a similar ``stretch'' to the data \citep{Line2021}. Specifically, we can write the observed flux, post-PCA and median division, as:

\begin{equation}
    \alpha\frac{P(\lambda,t)}{S(\lambda)}[1+\delta f(\lambda,t)]
\end{equation}

\begin{figure*}
    \centering
    \includegraphics[width=1.0\linewidth]{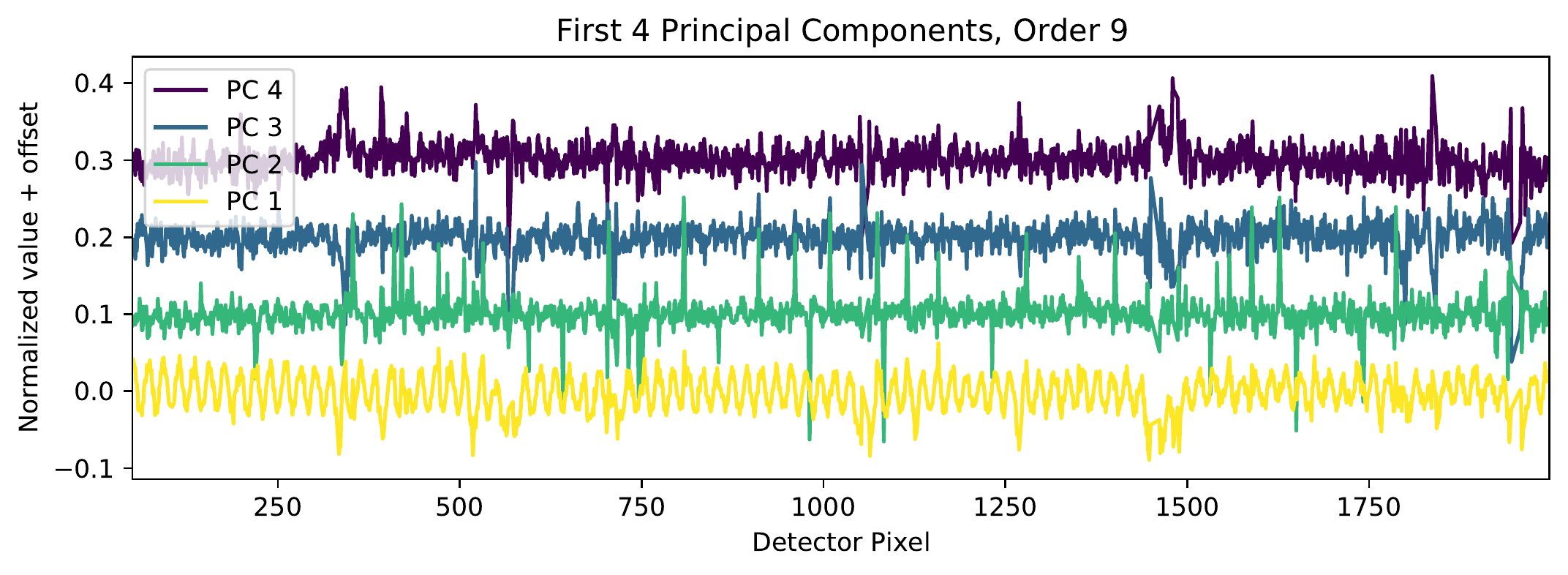}
    \caption{Example of the omitted principal components for the 2.44--2.49 $\mu$m NIRSPEC order shown in Figure \ref{fig:reduction}. The impact of time-varying fringing and telluric absorption is clear throughout the first 4 principal components. A telluric feature can be seen near pixel 1500, where the PCA is correcting variations from our baseline PSG model. Amplitudes are relative to the continuum-normalized spectrum, with a 0.1 offset added to successive principal components.}
    \label{fig:PCA}
\end{figure*}

Where we have subtracted the bias term from Equation 5. $\delta f(\lambda, t)$ represents the dropped principal components, which we use to ``stretch'' the model in our log-likelihood calculations and reproduce the $\delta f(\lambda, t)P(\lambda,t)/S(\lambda,t)$ term. Note that this assumes $\delta f(\lambda, t)$ does not contain any contribution from $P(\lambda, t)$, which is generally not valid, although the contribution may be negligible in some cases. These processing steps may inadvertently impact the strength of the planet features relative to the continuum. We therefore include a multiplicative scaling to the planet model ($\alpha$) as a free parameter in our retrievals, which we expect to be of order unity. The third panel of Figure \ref{fig:reduction} shows the post-PCA time series for one order, with the time-varying fringing effectively removed. We show an example of the dropped principal components in Figure \ref{fig:PCA}, which appear to be dominated by fringing and airmass variation. 

Occasionally the post-PCA data show a temporally fixed fringing pattern at much lower amplitude than the original fringe. While Figure \ref{fig:reduction} indicates this is not substantially impacting these observations, we include a second median division after PCA in order to eliminate any residual fringing, though doing so does not appear to impact the results of our retrievals. The fourth panel of Figure \ref{fig:reduction} shows the time series after this final division and an additional masking of values differing from the median by $>4\%$ (approximately 6 times the median absolute deviation)

\subsubsection{Negative Injection and Number of PCs}

Our initial tests found that removing 4 PCs enabled a detection of WASP-33b at a velocity similar to previous results from the literature \citep[e.g.][]{nugroho2021, vansluijs2022}. However, PCA will remove or distort the underlying planet signal, which must be mitigated in order to run retrievals. We adopt a negative-injection approach, where we subtract the proposed planet/star time series from the observed prior to performing the PCA. In the case where the planet model accurately reproduces the data, this eliminates the PCA self-subtraction issue, while potentially exaggerating it for poorly-matched models. This approach is sensitive to the exact match between the observations and the template, requiring PCA to be performed as part of each logL evaluation. While this significantly increases the computational cost of each log-likelihood call, it should reduce the sensitivity of the retrieval to the precise number of omitted principal components.

\begin{figure}
    \centering
    \includegraphics[width=1.0\columnwidth]{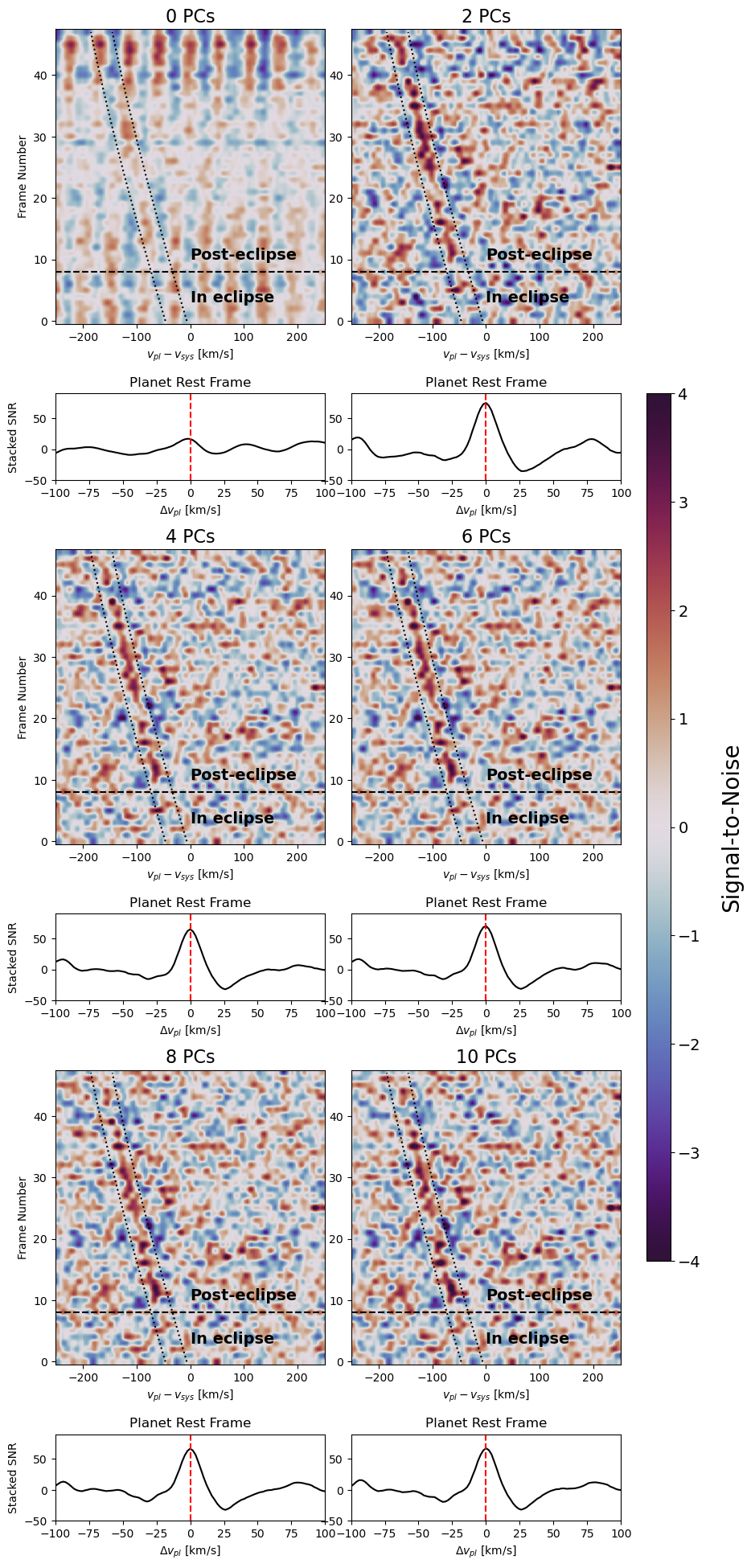}
    \caption{Cross-correlation versus planet velocity and exposure number for varying numbers of removed principal components. Data processing was identical as for the retrievals at the best-fit planet velocity, including negative injection before PCA. Signal-to-noise is estimated by dividing by the variance for $v_{pl} - v_{sys} > 0$, which does not overlap with the planet track. Lower panels show the summed CCF in the planet rest frame. Without PCA, the cross-correlation space is dominated by the time-varying fringe associated with transmissive optics in KPIC. PCA almost completely eliminates this effect, enabling a clear detection of the planet signal. Removing between 2 and 10 PCs has a negligible impact on the planet detection. We choose to remove 4 PCs, as visual inspection shows significant fringing and telluric contributions even in the 3rd and 4th components (see Figure \ref{fig:PCA}).}
    \label{fig:pc_track}
\end{figure}

Using PCA requires a decision as to the number of principal components to omit. Removing too few principal components can lead to significant contamination from fringing or tellurics, while removing too many can distort or even eliminate the planet signal. We believe that our negative-injection approach should substantially reduce the sensitivity of the planet detection to the number of omitted PCs, provided the planet is injected at approximately the correct strength relative to the continuum. 

To test the impact of varying the number of omitted PCs, we computed the stellar-frame cross-correlation function of the maximum-likelihood retrieved planet spectrum with each frame in the 21 November data for varying numbers of omitted PCs. This is shown in Figure \ref{fig:pc_track}, where the CCF has been divided by the variance for $v_{pl} - v_{sys} > 0$ in order to estimate the signal-to-noise ratio. We then shift to the planet rest frame and sum in order to estimate a 1D CCF. Without PCA, the 2D cross-correlation space is clearly dominated by the time-varying fringe, and the planet signal is barely detectable even in the stacked CCF. Between 2 and 10 omitted principal components, the planet signal is clear even in the 2D CCF, and the number of PCs dropped does not appear to significantly change the CCF.

The apparent independence of the planet detection on the number of PCs makes the choice of how many PCs to omit somewhat arbitrary. We opted to omit 4 PCs based on examination of the components in Figure \ref{fig:PCA}, which show substantial contributions from fringing and tellurics in these components. While Figure \ref{fig:pc_track} suggests we could safely drop additional components, we are concerned that model mismatch may still result in self-subtraction when many PCs are dropped, despite the negative injection. Omitting 4 components is a qualitative balance between removing known contaminants while avoiding self-subtraction issues and is consistent with other high-resolution retrievals \citep[e.g][]{Line2021}. For consistency and to avoid possible overfitting, we omit the same number of components for and all orders. In the future, it would be preferable to develop and use a quantitative set of criteria for the number of PCs to remove. Alternatively, improved forward modeling of the KPIC fringing could reduce or eliminate the need for PCA. 

\subsubsection{Template Spectra and Opacities}\label{ssec:templates}
Calculating a log-likelihood requires comparing the detrended time series obtained from the steps described in the previous subsection to an analogous simulated data set with known orbital and planetary parameters. This requires models of both stellar and planetary spectra in order to calculate the equivalent to the $P(\lambda, t)/S(\lambda)$ term obtained above.

For the stellar template, we use a PHOENIX model \citep{phoenix} with $\rm T_{eq} = 7400\ K$, $\rm [Fe/H] = 0.0$, and $\log g = 4.5$, chosen to be similar to the stellar parameters in Table \ref{tab:props}. We then simulate the observed stellar disk as a 60$\times$60 array of these 1D spectra, Doppler shift by the expected $v\sin i$ at the location of each grid point on the stellar disk, and apply limb darkening coefficients from \citet{sing2010}. We neglect the impacts of both gravity darkening and the $\delta$ Scuti pulsations of WASP-33. Each of these may lead to an achromatic systematic offset between our simulated $S(\lambda)$ and the true stellar spectrum, which the scaling parameter in the retrieval should account for. We note that \citet{cauley2021} reported the pulsation of WASP-33 leads to time-varying velocity shifts in the cross-correlation function for high resolution optical data. We do not expect this phenomenon to impact $K$-band observations, as we do not anticipate H$_2$O or CO features in the A-type stellar spectrum.

For the planet templates, we use \petit\ \citep{prt:2019, prt:2020} to perform the radiative-transfer calculation with 80 log-uniform spaced pressure slabs from $10^2$ bar to $10^{-6}$ bar and a spectral resolution of $1.25\times10^5$ over the $\sim2.1$--$2.5\ \rm \mu m$ range covered by our observations. We use molecular opacities for CO, H$_2$O, and OH generated using ExoCross \citep{exocross2018} from the HITEMP linelists \citep[HITEMP 2010 for H$_2$O, HITEMP 2019 for CO, HITEMP 2020 for OH;][]{hitemp2010, hitemp2020}. We use the \citet{li2015} partition function for CO, the \citet{polyansky2018} partition function for H$_2$O, and the \citet{YOUSEFI2018} partition function for OH. For SiO, we use the ExoMol-recommended linelists and partition functions from \citet{Yurchenko2021}. Opacities are generated on pressure-temperature grids ranging from $10^3$--$10^{-6}$ bar in pressure and from 80--6000 K in temperature, which \petit\ interpolates to the desired pressure/temperature values for each layer. We allow molecular abundances for each species to vary freely during the retrieval but hold abundances constant at all pressures.

\subsubsection{Pressure-Temperature Profile}\label{ssec:ptcloud}

To describe the atmospheric temperature at each pressure layer, we use the pressure-temperature (PT) profile parameterization from \citet{madhusudhan2009}. This 6-parameter model divides the atmosphere into three zones --- two upper zones with exponential behavior and an isothermal lower atmosphere. Using two upper zones allows this model to fit both inverted and non-inverted PT profiles, and the isothermal lower atmosphere is expected for hot Jupiters due to high optical depths \citep{madhusudhan2009}. Based on previous results from WASP-33b \citep{nugroho2021, vansluijs2022, yan2022} we require a thermal inversion for the dayside PT profile. This requirement speeds convergence and avoids issues with fitting inverted sidelobes of the cross-correlation function. We also enforce $\rm T < 6000 K$ throughout the atmosphere in order to avoid hitting the limits of our opacity tables.

\subsubsection{H$^-$}\label{ssec:hminus}

The temperature and irradiation of WASP-33b suggest a significant fraction of upper atmospheric H$_2$ may be dissociated, and H$^-$ may be a significant source of continuum opacity. We experimented with several ways of incorporating H$^-$. Our first approach used the \texttt{poor\_mans\_nonequ\_chemistry} calculator from \petit\ \citep{molliere2017, prt:2020} to estimate the chemical equilibrium H$_2$, He, H, H$^-$, and e$^-$ fractions in the atmosphere for the proposed PT profile, which \petit\ then uses to calculate H$^-$ opacity. While this accounts for the strongly-varying vertical abundance of H$^-$, this is inconsistent with our free-retrieval approach used for other species. We next attempted to include H$^-$ and e$^-$ as species in the retrieval, but both were effectively unconstrained over a prior based on the typical equilibrium chemistry values.

\begin{figure*}
    \centering
    \includegraphics{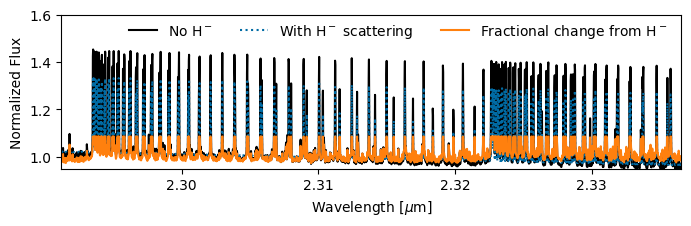}
    \caption{Normalized planet spectrum including H$^-$ opacity (dotted blue) and without (black) over the wavelength range of a single NIRSPEC order corresponding to the CO bandhead. The ratio of the spectrum without H$^-$ to the spectrum with H$^-$ is plotted in orange. Over the short bandpasses of the NIRSPEC orders, H$^-$ is effectively achromatic, reducing line strengths by $\sim10$\% relative to the continuum. As the continuum is removed during the detrending, the overall scaling parameter should account for the impact of H$^-$, explaining the weak constraints using a free retrieval approach and justifying the omission of H$^-$ opacity.}
    \label{fig:Hminus}
\end{figure*}

Finally, after comparing model planet spectra with and without H$^-$ (see Figure \ref{fig:Hminus}), we opted not to include H$^-$ opacity in our retrievals. We find that on the scale of a single NIRSPEC order, H$^-$ is effectively achromatic, resulting in a consistent $\sim10\%$ reduction in line strengths (see Figure \ref{fig:Hminus}, orange line) relative to the continuum. As we do not preserve the continuum level through the detrending process, the overall scaling parameter is already effectively scaling the line strengths relative to the continuum, making the additional inclusion of H$^-$ opacity redundant. The CO and H$_2$O abundances retrieved without H$^-$ opacity are statistically consistent with the retrievals including H$^-$ based on equilibrium abundances.

Neglecting H$^-$ should result in a smaller value for the scaling parameter, as the lines will be weaker relative to the continuum than we assume. This achromatic behavior may not hold over wider bandpasses, in which case H$^-$ should be included in the retrieval model.

\subsubsection{Model Processing}\label{ssec:modproc}

Using the planet spectra calculated by \petit, we next calculate a model $P(\lambda, t)/S(\lambda)$ time series. We first scale the planet flux by the planet/star area ratio and then Doppler shift the planet to the proposed velocity for each exposure. We subsequently divide the planet model by the stellar model, Doppler shift by the systemic velocity, and interpolate onto the observed wavelength grid. We then convolve the model spectrum with a 1.7 pixel Gaussian representing the NIRSPEC instrument profile and divide by the time-series median. The optimum kernel was determined by maximizing the log-likelihood at the planet peak identified in a $K_p-v_{sys}$ diagram with a fixed planet template and is slightly larger than expected based on the instrument resolution, most likely as a result of the motion of WASP-33b along its orbit within a single set of exposures(known as Doppler smearing).  Finally, we apply the PCA ``stretch'' to the model and perform a second median division in order to match the treatment of the observations. At this point the model spectra should be exactly analogous to the data, and we calculate the log-likelihood as described in \citet{brogi2019}. 

\subsubsection{Parameter Selection}\label{ssec:pars}
\begin{deluxetable*}{ccc}
    \tablehead{\colhead{Name} & Symbol  & \colhead{Prior}}
    \startdata
        Upper-atmosphere scaling & $\alpha_1$ &  Uniform(0,1) \\
        Lower-atmosphere scaling & $\alpha_2$  &  Uniform(0,1) \\
        Top-of-atmosphere temperature [K] & T & Uniform(2000, 6000) \\
        Reference pressure [bar] & log P$_1$ & Uniform(-6, 0) \\
        Mid-atmosphere fraction & $f_{\rm P_2}$ &  Uniform(0,1) \\
        Lower-atmosphere fraction &  $f_{\rm P_3}$  & Uniform(0,1)\\
        $K_p$ offset [\kms] & $\Delta K_p$  & Uniform(-40, 40)  \\
        $v_{sys}$ offset [\kms] & $\Delta v_{sys}$ & Uniform(-30, 30)  \\
        log H$_2$O mass-mixing ratio & log H$_2$O  &  Uniform(-6, -1) \\
        log CO mass-mixing ratio & log CO &  Uniform(-6, -0.1) \\
        log OH mass-mixing ratio & log OH  & Uniform(-6, -1) \\
        log SiO mass-mixing ratio & log SiO &  Uniform(-6, -1) \\
        log $\rm ^{13}CO/^{12}CO$ & $\rm log ^{13}CO_{rat}$ &  Uniform(-2.5, -0.5) \\
        Total H mass-fraction & log allH &  Uniform(-0.2, -0.05) \\
        HI fraction & $f_{\rm HI}$  & Uniform(0, 1) \\
        Scale factor & scale & Uniform(0, 2) \\
    \enddata 
    \caption{List of parameters and priors for the WASP-33b atmospheric retrievals. In addition to these priors, we required both a thermal inversion and that the atmospheric temperature stay below 6000 K at all pressure levels}, in order to avoid going beyond the bounds of our opacity grids.
    \label{tab:priors}
\end{deluxetable*}

Our retrieval model consists of 16 free parameters. We list these parameters and priors in Table \ref{tab:priors}, and provide a brief description and motivation here. For the PT profile, we adopt the six-parameter model from \citet{madhusudhan2009}, setting $\beta = 0.5$. This takes a top-of-atmosphere temperature, two coefficients to scale the argument of the exponential functions, and three reference pressures for the transition between the different exponentials and the isothermal lower atmosphere. We implement this as a single physical pressure, and two fractional breakpoints (in log space) between that pressure and the top/bottom of the atmosphere respectively. This allows inequality constraints to be more easily placed on the different pressures in order to enforce the presence/absence of a thermal inversion. We choose priors for the temperature based on previous literature values \citep{nugroho2021, vansluijs2022}, while we allow the scaling and pressure parameters to vary over a broad range. We require a thermal inversion based on the initial qualitative cross-correlation analysis discussed in Section \ref{sec:res} and previous literature \citep{nugroho2021, yan2022, vansluijs2022}.

We also fit two parameters for the orbital properties of WASP-33b. While the planet radial velocity semi-amplitude ($K_p$) and systemic velocity ($v_\mathrm{sys}$) are well-constrained by previous literature, the 3D atmospheric structure may introduce phase-dependent velocity shifts due to winds or outflows \citep{beltz2022}. We choose priors based on the approximate extent of the planet peaks obtained from cross-correlation with an assumed template (see Figure \ref{fig:kpvsys}) and to cover the range of values reported in \citet{vansluijs2022}.

We fit for seven abundances -- H$_2$O, CO, OH, SiO, H$_2$, the $^{13}$CO/$^{12}$CO ratio, and the H$_2$ dissociation fraction. H$_2$O and CO are expected to dominate the $K$-band spectrum for UHJs and are expected to be the dominant forms of C, O, and overall metals in the atmosphere of WASP-33b. We also include $\rm ^{13}CO$, as we expect it to be detectable in an atmosphere with a high overall CO abundance \citep{molliere2019}. While simulations suggested that the limited wavelength coverage of our observations may be insufficient to make an  independent detection of OH, we include it for completeness based on the previous dayside detection reported by \citet{nugroho2021}, which used prominent OH features in the $H$-band. \citet{brogi2023} also detected OH in the atmosphere of the UHJ WASP-18b with $H$-band observations. We also fit for SiO, as the \petit~equilibrium chemistry models suggest that it may be a significant oxygen reservoir in the high-C/O, high-metallicity regime, and \citep{cont2021} previously reported a detection of atomic Si in the dayside upper atmosphere of WASP-33b. The final abundance parameters are the total (molecular and atomic) hydrogen mass fraction and the H$_2$ dissociation fraction. The remainder of the atmosphere is assumed to be entirely He.

Finally, we fit an additional multiplicative scaling parameter applied to the planet model. This accounts for a number of possible biases in our retrieval setup, while also serving as a check for their severity. Ideally, we should retrieve a scaling parameter of 1. However, we neglected both $\delta$ Scuti pulsations and gravity darkening of the host star, which may introduce systematic differences in the relative flux of the planet. We also ignore H$^-$ opacity in the planet atmosphere, which could produce achromatic changes in the line strengths. The scale factor allows us to correct for these omissions, while also checking the size of these effects. 

\subsubsection{Nested Sampling}

We compare our forward-models described above to the processed observations using the log-likelihood function from \citet{brogi2019}. Specifically:

\begin{equation}
    \log L = -\frac{N}{2}\log\left(\frac{D^2}{N}+\frac{M^2}{N}-2\frac{M*D}{N}\right)
\end{equation}

Where $N$ is the number of valid points in each observed spectrum, $D$ is the cleaned data, and $M$ is the proposed spectrum, which already includes the scaling parameter. This is calculated for each order and each science frame taken out-of-eclipse, then summed to produce a total log-likelihood for the observation sequence. 

This log-likelihood calculation enables us to use the full range of Bayesian fitting tools. As in \citet{Line2021}, we opt for a nested-sampling approach \citep{skilling2004}, in our case the \dynesty\ implementation \citep{speagle2020}. Nested sampling is suitable for high-dimensional problems with computationally expensive likelihood functions and potentially complicated posteriors, making it ideal for atmospheric retrievals. It is also easily parallelizable, allowing us to take advantage of cluster computing resources. Our retrievals using 16 CPU slots typically take $\sim$2--4 days to reach $\Delta\log z = 0.001$, which is sufficient to obtain good estimates of the posteriors.

\subsection{Validation Simulations}\label{ssec:test}
\begin{deluxetable}{ccc}
    \tablehead{Symbol & \colhead{Input} & \colhead{Retrieved}}
    \startdata
        $\alpha_1$ & 0.48 & $0.5^{+0.3}_{-0.3}$ \\
        $\alpha_2$ & 0.10 & $0.2^{+0.1}_{-0.1}$ \\
        T & 3420 &  $3760^{+740}_{-700}$ \\
        log P$_1$ & -2.6 &  $-2.7^{+0.8}_{-0.7}$ \\
        $f_{\rm P_2}$ & 0.29 & $0.2^{+0.2}_{-0.1}$ \\
        $f_{\rm P_3}$ & 0.42 & $0.2^{+0.2}_{-0.1}$ \\
        $\Delta K_p$ [\kms] & 0 & $1.8^{+3.3}_{-3.4}$ \\
        $\Delta v_\mathrm{sys}$ [\kms] & 0 & $0.8^{+1.6}_{-1.7}$ \\
        log H$_2$O & -4.3 & $-4.1^{+0.6}_{-0.7}$ \\
        log CO & -1.5 & $-1.4^{+0.5}_{-0.8}$ \\
        log OH & -2.5 & $-4.2^{+1.2}_{-2.0}$ \\
        log SiO & -3.7 & $-3.1^{+1.6}_{-2.0}$ \\
        log $\rm ^{13}CO_{rat}$ & -1.8 & $-2.1^{+0.3}_{-0.3}$ \\
        log allH & -0.13 & $-0.13^{+0.05}_{-0.05}$ \\
        log $f_{\rm HI}$ & 0.42 & $0.5^{+0.3}_{-0.4}$ \\
        scale & 1.0 & $1.2^{+0.6}_{-0.6}$ \\
        $\rm SNR_{pt}$ & 110 
    \enddata 
    \caption{Input and retrieved parameters for the test retrieval. Orbital phase sampling was identical to the 2021 November 21 data. The test case was chosen to be similar to the retrieval results, so that it serves as an injection/recovery test. }
    \label{tab:simulated}
\end{deluxetable}

As our retrieval framework by necessity includes a forward-modeling capability, we can easily test our retrievals on a simulated system with known inputs. This provides a check on systematic biases in the retrieval and ensures that the reported uncertainties are reasonably accurate. These checks also allow us to determine if any parameters are inherently poorly constrained by the available data. All simulated data sets used the same orbital phase sampling as the observations (2021 November 21). 

As our forward modeling does not include fringing or airmass variation, we do not include the PCA step in the test retrievals. The planet spectrum is the only time-varying component in the simulated spectrum. Even with our negative-injection approach, the planet spectrum is distorted by the PCA in the absence of stronger fringe/airmass-induced variations. This limits our ability to identify biases introduced from the PCA and may result in a mild overestimation of the retrieved errors in the simulations, as the negative-injection PCA approach is expected to increase preference for the true planet model. Ongoing developments in forward modeling of the KPIC fringing should reduce future reliance on PCA for defringing.

Our first test case was a star-only (no planet) simulated data set. This provides a non-detection baseline for comparison. Figure \ref{fig:nocorner} shows the full corner plot \citep{corner} in Appendix \ref{app:corner}. Despite using the same stopping criteria as other retrievals, the star-only simulation shows much flatter and noisier posteriors and reached the stopping criteria in far fewer iterations. Additionally, the scale parameter shows a weak preference towards smaller values, suggesting a preference for the absence of the planet, as expected. The retrieved pressure-temperature profile is nearly isothermal, and the maximum-likelihood planet spectrum is nearly flat, as the fitting minimizes the strength of the planet features.

Our second test case was based on the retrieved parameters for WASP-33b listed in the ``Input'' column of Table \ref{tab:simulated}. The signal-to-noise ratio of the simulations was chosen to be similar to that of the observations. This provides both a general check for systematics or unconstrained parameters in our retrieval framework as well as a specific injection/recovery test for the retrieved dayside atmosphere. Retrieved values and uncertainties are listed in the ``Retrieved'' column of Table \ref{tab:simulated}, and the full corner plot is presented in Figure \ref{fig:corner_test1}. 

All parameters except the OH mass fraction are retrieved to within 1$\sigma$ of the input values, with error bars similar to those retrieved from the observations. The PT profile retrieved from the simulations is slightly hotter than the input profile but is generally within the 1$\sigma$ bound and does not impact the retrieved abundances. The SiO mass fraction, H mass fraction, and H dissociation fraction are all poorly-constrained in both the observations and the simulated retrieval, suggesting our observations have limited sensitivity to these parameters. 

While the OH posterior shows a peak in the observations, this is not the case in the simulated retrieval. The $^{13}$CO isotopologue ratio and the scale factor are also better-constrained in the observations than the simulations, though not to the same degree as the OH abundance. While this may be a result of the negative-injection PCA approach used for the observations improving sensitivity to relatively small changes in the spectrum, the poor sensitivity in simulated retrievals suggests both apparent constraints may be artifacts.

We also ran retrievals using \texttt{emcee} \citep{emcee}. This provides a check that any biases are not due to the particulars of our nested sampling approach, such as the stopping criteria or the number of live points. The resulting median values and uncertainties from \texttt{emcee} were consistent with the nested sampling results. This suggests that systematics are more likely to be related to limitations in the radiative transfer calculation, log-likelihood function, or inherent limitations in the data, rather than the statistical framework used.

\section{Results} \label{sec:res}
We present qualitative results from cross-correlation analysis in Section \ref{ssec:cc}. Quantitative results from our retrieval framework are presented in Section \ref{ssec:day}.

\begin{figure}
    \centering
    \includegraphics[width=1.0\linewidth]{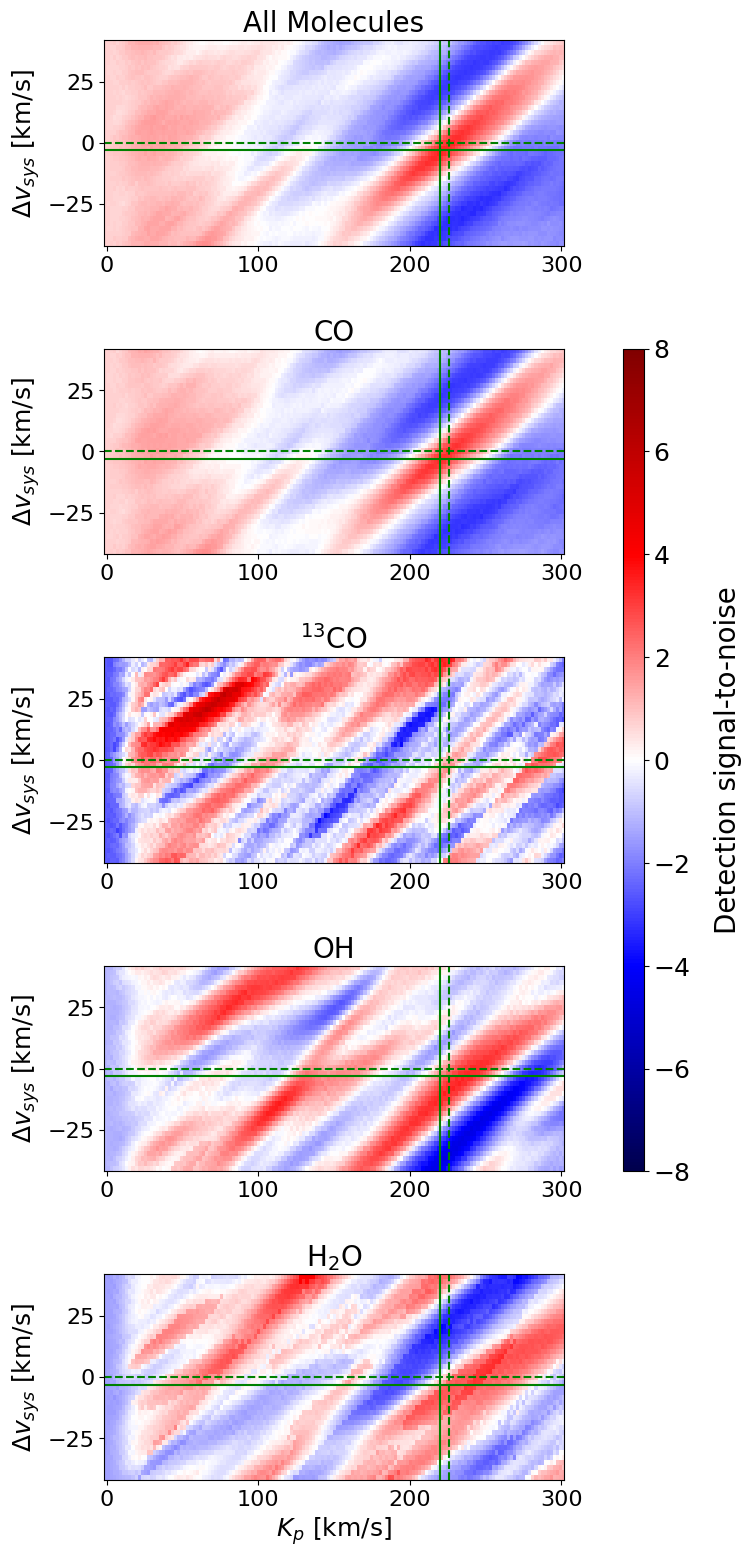}
    \caption{$\Delta v_\mathrm{sys} - \Delta K_p$ diagrams for the retrieved planet spectrum. The nominal and retrieved velocities are indicated in dashed and solid green, respectively. The top row shows all molecules, while the subsequent four rows show the contribution of individual species. $\rm ^{12}CO$ dominates the detection, while the other species show relatively weak features at the expected velocity of WASP-33b. Detection strength is estimated by dividing the computed log-likelihood map by the standard deviation taken after masking $K_p < 50$ \kms and below/above the 10th/90th percentiles in order to reduce the impact of the PCA-induced feature near $K_p = 0$ \kms and the planet peak itself. Figure \ref{fig:pc_track} provides a better quantitative estimate of detection strength.}
    \label{fig:kpvsys}
\end{figure}

\begin{deluxetable}{ccc}
    \tablehead{\colhead{Name} & Symbol & \colhead{Retrieved}}
    \startdata
        Upper-atmosphere scaling & $\alpha_1$  & $0.6^{+0.3}_{-0.3}$ \\
        Lower-atmosphere scaling & $\alpha_2$ &  $0.1^{+0.07}_{-0.05}$ \\
        Top-of-atmosphere temperature [K] & T & $3500^{+500}_{-500}$ \\
        Reference pressure [bar] & log P$_1$  & $-2.8^{+0.7}_{-0.6}$ \\
        Mid-atmosphere fraction & $f_{\rm P_2}$  & $0.2^{+0.2}_{-0.1}$ \\
        Lower-atmosphere fraction &  $f_{\rm P_3}$ &  $0.5^{+0.3}_{-0.3}$ \\
        $K_p$ offset [\kms] & $\Delta K_p$  &  $-6^{+4}_{-3}$ \\
        $v_{sys}$ offset [\kms] & $\Delta v_{sys}$ & $-3^{+2}_{-1}$  \\
        log H$_2$O mass-mixing ratio & log H$_2$O &  $-4.1^{+0.7}_{-0.9}$ \\
        log CO mass-mixing ratio & log CO  & $-1.1^{+0.4}_{-0.6}$ \\
        log OH mass-mixing ratio & log OH  & $-2.1^{+0.5}_{-1.1}$\\
        log SiO mass-mixing ratio & log SiO  & $-3.6^{+1.3}_{-1.5}$ \\
        log $\rm ^{13}CO/^{12}CO$ & log $\rm ^{13}CO_{rat}$ &  $-1.7^{+0.3}_{-0.5}$ \\
        Total H mass fraction & log allH   & $-0.14^{+0.06}_{-0.04}$ \\
        HI fraction & $f_{\rm HI}$ & $0.5^{+0.3}_{-0.3}$ \\
        Scale factor & scale  & $1.3^{+0.5}_{-0.4}$ \\
        \hline
        Derived Parameters \\ 
        C/O ratio & C/O  & $0.8^{+0.1}_{-0.2}$ \\
        log C/H$\rm _{VMR}$ & log C/H  & $-2.4^{+0.4}_{-0.6}$ \\
        log O/H$\rm _{VMR}$ & log O/H  & $-2.3^{+0.4}_{-0.5}$\\
    \enddata 
    \caption{Retrieval results for of WASP-33b. We report the median and $\pm34\%$ quantiles for all parameters, which were in good agreement with the corresponding values obtained from an MCMC. Full corner plots are included in Appendix \ref{app:corner}. In several cases, the peak of the retrieved posteriors shows substantial offsets from the median of the distribution, which can be seen in the full corner plots. Strong covariances between abundance parameters enables a better constraint on the C/O ratio than would be expected from the marginalized posteriors alone (see Figure \ref{fig:CO_CH}).}
    \label{tab:res}
\end{deluxetable}

\subsection{Cross-Correlation Analysis}\label{ssec:cc}

We use the best-fit retrieval parameters to compute $\Delta v_\mathrm{sys}-\Delta K_p$ diagrams, both for the overall model and for each individual species. These are plotted in Figure \ref{fig:kpvsys}, with the retrieved $\Delta v_\mathrm{sys}$ and $\Delta K_p$ values indicated in dashed green. This is the approach that previous high-resolution studies have used to make molecular detections \citep[e.g.][and many others]{brogi2012, buzard2020}. We note that attempts to quantify strengths for this type of detection have been fraught \citep[e.g.][]{buzard2020, Finnerty2021, buzard2021}, but these plots can still provide useful qualitative insight.

Specifically, quantitative estimates of the detection strength from $K_p-v_{sys}$ plots typically assume values far from the planet peak are normally-distributed and use the standard deviation far from the planet as an estimate of the noise. In the case of Figure \ref{fig:kpvsys}, this is complicated by the clear presence of systematic variations as a function of $K_p$ due to the detrending suppressing the planet at small $K_p$. We therefore estimate the noise in Figure \ref{fig:kpvsys} by first masking $K_p < 50$ \kms to exclude the region where the PCA is having a significant impact, then additionally masking values above/below the 10th/90th percentile to remove the planet feature. We then calculate the standard deviation and use this to make the detection strength map. These thresholds are arbitrary, and the remaining points after masking still do not appear to be normally distributed. We emphasize that Figure \ref{fig:kpvsys} is intended for a qualitative assessment of the detection of different species. For a quantitative assessment, Figure \ref{fig:pc_track} is better for estimating the strength of the overall detection, and Table \ref{tab:bayes} presents Bayes factors for the detection of each molecules.

The strong dependence of the detrending signal on $K_p$ allows it to be removed by subtracting each column of Figure \ref{fig:kpvsys} by its median. This results in a detection strength of 0 at $K_p = 0$, as expected when the planet is totally removed by detrending. For the all molecule case, the planet is detected at an SNR$\sim$12, which is more consistent with expectations based on Figure \ref{fig:pc_track} than the scale of Figure \ref{fig:kpvsys}. However, this removal of the detrending systematics is not statistically rigorous. We emphasize that Figure \ref{fig:pc_track} offers the most robust estimate of the overall detection strength.

The planet is clearly detected in the $K_p-v_{sys}$ diagram plotted in the top panel of Figure \ref{fig:kpvsys}. In the 2D plots of cross-correlation versus time in Figure \ref{fig:pc_track}, planet velocity track is clearly visible, and the retrieved velocities are in good agreement with the nominal values. Additional validation of the detection is provided by the disappearance of the planet feature during secondary eclipse in Figure \ref{fig:pc_track}.

The cross-correlation detection is dominated by $^{12}\rm CO$ based on comparing the first and second panels of Figure \ref{fig:kpvsys}. OH also produces a strong signal at the expected location in the $K_p-v_{sys}$ space (fourth panel of Figure \ref{fig:kpvsys}), while the H$_2$O template produces a somewhat weaker signal (fifth panel of Figure \ref{fig:kpvsys}). The $^{13}\rm CO$ template has a weak feature coincident with the expected planet velocity, but this feature does not dominate and we do not consider this to be an independent detection of $^{13}\rm CO$. The SiO template does not produce any features in the $K_p-v_{sys}$ space and is not shown.

\subsection{Retrieval Results}\label{ssec:day}
\begin{figure*}
    \centering
    \includegraphics[width=2.1\columnwidth]{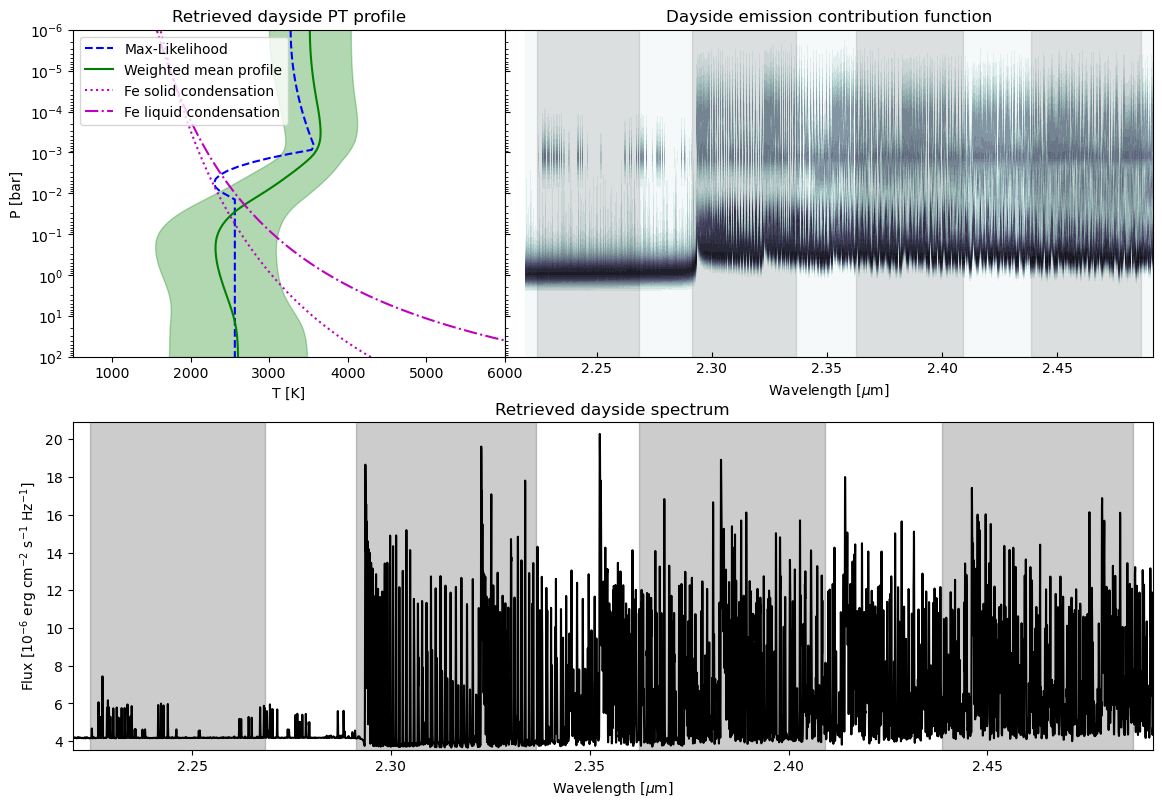}
    \caption{Retrieved PT profile (top left), $K$-band emission contribution (top right), and spectrum (bottom) for the best-fit atmospheric parameters. The weighted mean of all PT profiles is plotted in solid green, with the 1$\sigma$ range shaded. The PT profile from the maximum-likelihood parameters is plotted in dashed blue and generally agrees well, though with a somewhat shallower and higher inversion. Condensation curves for solid iron and liquid iron from \citet{ackerman2001} are overplotted, though we do not expect dayside clouds. In the right panel, the contribution of each pressure layer to the overall planet emission with the best-fit abundances and median PT profile is indicated by the shading, with the observed NIRSPEC orders overplotted in grey. The observed emission emerges primarily from the 1-0.1 bar range. The CO line cores beyond $\sim2.3$ $\mu$m include significant contributions from the upper atmosphere. The bottom panel shows the spectrum obtained from the median parameters convolved to the approximate resolution of NIRSPEC, again with the NIRSPEC orders overplotted in grey.}
    \label{fig:dayspec}
\end{figure*}

Retrieved atmospheric parameters are listed in Table \ref{tab:res}. The full corner plot \citep{corner} for the retrieval is presented in Figure \ref{fig:daycorner}. The retrieved dayside PT profile, emission contribution function, and median spectrum are plotted in Figure \ref{fig:dayspec}. 

The dayside atmosphere shows a clear inversion beginning at approximately 0.1 bar, consistent with previous results \citep[e.g.][]{nugroho2021, vansluijs2022}. The $K$-band emission contribution function (Figure \ref{fig:dayspec}, top right panel) indicates that the bulk of the observed emission arises near this pressure, though CO continues to contribute to the line cores from altitudes up to $\sim3\ \mu$bar. Pressures greater than $\sim1$ bar do not contribute to the observed emission, and the retrieved PT profile is therefore unlikely to be accurate at these pressures. We note that the high-resolution spectroscopy is sensitive to the line strength relative to the continuum, which is determined by the shape and contrast of the PT profile, the atmospheric metallicity, the scaling parameter, and possible impacts from H$^-$. This can allow ``cold'' PT profiles to produce similar output spectra to those of ``hotter'' PT profiles. While temperature will also change the relative strengths of different lines, this is a comparatively small effect over a limited bandpass. We therefore caution against relying on the exact temperatures of the retrieved PT profiles.

We obtain log mass fractions of $-4.1^{+0.7}_{-0.9}$ for H$_2$O, $-1.1^{+0.4}_{-0.6}$ for CO, and $-2.1^{+0.5}_{-1.1}$  for OH. This indicates that H$_2$O is almost entirely dissociated on the dayside of WASP-33b, consistent with previous findings from \citet{nugroho2021}. SiO is effectively unconstrained due to its relatively low $K$-band opacity, but peaks at a log mass fraction of $-3.6$, which would be consistent with the neutral Si detection reported in \citet{cont2022}. We also obtain a weak constraint on the $^{13}$CO/$^{12}$CO ratio, peaking around $10^{-1.6}$ but with a significant tail giving a median value of $10^{-1.7}$. The abundance parameters show a strong positive covariance, which enables better constraints on the C/O ratio than would be expected from the marginalized uncertainties. 

Finally, we note that while the scaling parameter peaks near unity, there is a long tail towards higher values. The scaling parameter controls the strength of lines relative to the host star continuum, which can also be influenced by changes in the PT profile. Figure \ref{fig:daycorner} shows weak degeneracies between the PT profile parameters and the scaling parameter, with larger values of the scaling parameter corresponding to PT profiles which produce weaker lines. This suggests the tail of the scaling parameter is a result of under-constraining the PT profile.

There are a number of systematics or physical effects that could also change the value of the scaling parameter. Our test retrievals suggest this may be a result of our data processing (see Figure \ref{fig:corner_test1}). Physical factors include underestimates of the planet radius or temperature, H$^-$ opacity, thermal expansion of the dayside atmosphere, extended emission from an outflow, or stellar pulsations. The retrieved posterior peaks at a scale factor $\sim1$, suggesting that these effects are either relatively minor or largely cancel. 

\section{Discussion} \label{sec:disc}
We present Bayes factors for the detection of each molecule compared with a flat planet model in Table \ref{tab:bayes} and discuss the detection significance for each molecule in Section \ref{ssec:detconf}. We discuss the retrieved wind speeds and thermal structure in Section \ref{ssec:daynight}. The C/O ratio and metallicity of WASP-33b are discussed in Section \ref{ssec:ctoo}, including a comparison to equilibrium chemistry models. We discuss the weak constraint retrieved for the CO isotopologue ratio in Section \ref{ssec:isotope} and finally discuss the implications of our results for the formation history of WASP-33b in Section \ref{ssec:formation}.

\subsection{Detection Confidence}\label{ssec:detconf}

\begin{deluxetable}{cc}
    \tablehead{\colhead{Model} & \colhead{Bayes Factor}}
    \startdata
        All molecules & 111  \\
        CO only & 102  \\
        $^{13}$CO only & 9.8 \\
        H$_2$O only & 6.9  \\
        OH only & 5.6  \\
    \enddata 
    \caption{Bayes factors comparing a flat planet model with the best-fit retrieved planet model and the best-fit model corresponding to each individual molecule. This allows us to assess the detection strength of each species.}
    \label{tab:bayes}
\end{deluxetable}

To assess the strength of our detections, we compute the Bayes factor comparing a flat/no planet model and the best-fit models for all molecules together and for each molecule independently. These values are presented in Table \ref{tab:bayes}. The best-fit planet model is strongly preferred ($\rm \Delta BIC > 100$) , with CO dominating the detection. The evidence for H$_2$O and OH is substantial but weaker than CO, consistent with expectations based on the weaker features in the $K_p-v_{sys}$ space in Figure \ref{fig:kpvsys}.

\subsection{Winds and Thermal Structure}\label{ssec:daynight}

Consistent with both previous results \citep{nugroho2021, vansluijs2022} and global circulation models of UHJs \citep{wardenier2021, komacek2022, beltz2022}, our retrievals confirm our assumption of a thermal inversion on the dayside. We find that constant offsets in the PT profile appear to have a limited impact on the final spectrum, suggesting our observations are not particularly sensitive to absolute temperature, with the curvature of the PT profile having a much more substantial impact. The inclusion of the scaling parameter may compound this, as it provides an alternative way to change line depths in the planet model. Figure \ref{fig:dayspec} shows that our $K$-band observations are most sensitive to a relatively narrow range of pressures and therefore have limited ability to constrain the overall PT profile. Additional observations at other bands may improve the vertical constraints.

The velocity parameters prefer a slight blueshift compared to the nominal values in Table \ref{tab:props}. While this could be a result of morning-to-evening winds, the shift is only 1-2$\sigma$ and is well within the model-dependent variations in velocity parameters reported in \citet{vansluijs2022}. Additional observations providing continuous coverage of a larger phase range would provide better constraints on dayside winds.

We adopted the $-9.2$ \kms systemic velocity from \citet{gontcharov2006}. More recent observations have found $v_{sys}$ in the range of -3 \kms \citep{collier2010} to 0 \kms \citep{nugroho2021}. \citet{vansluijs2022} found systemic velocities from -15 \kms to +6 \kms, depending on the orbital phase of the observations and the type of model used for cross-correlation. A more positive systemic velocity would imply larger blueshifts in the retrieved planet parameters relative to the expected Keplerian motion in the host star reference frame.

\subsection{C/O ratio and Metallicity}\label{ssec:ctoo}

\begin{figure}
    \centering
    \includegraphics[width=1.0\linewidth]{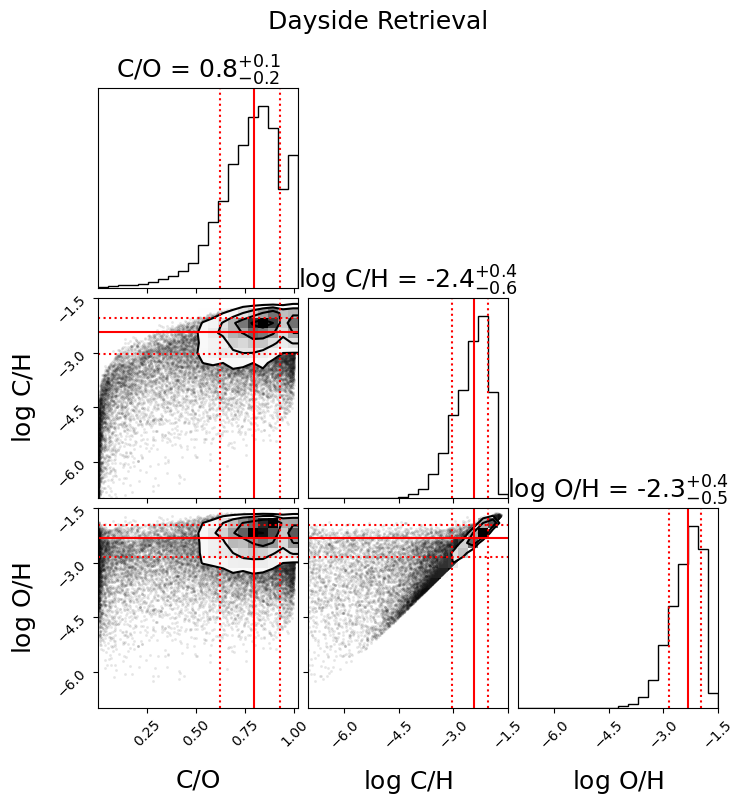}
    \caption{Retrieved C/O, C/H, and O/H number ratio posteriors. The C/O ratio is better constrained than the individual C/H and O/H ratios, consistent with our expectation that high-resolution spectroscopy over a narrow band is more sensitive to abundance ratios than absolute abundances due to the loss of continuum information during the data processing. The prior that the total mass fraction of all species be $<1$ can be seen in the absence of points to the upper left of the 2D plots, where low C/O ratio and high C abundances would require more than the entire remaining atmospheric mass in the form of H$_2$O.}
    \label{fig:CO_CH}
\end{figure}

\begin{figure}
    \centering
    \includegraphics[width=0.8\linewidth]{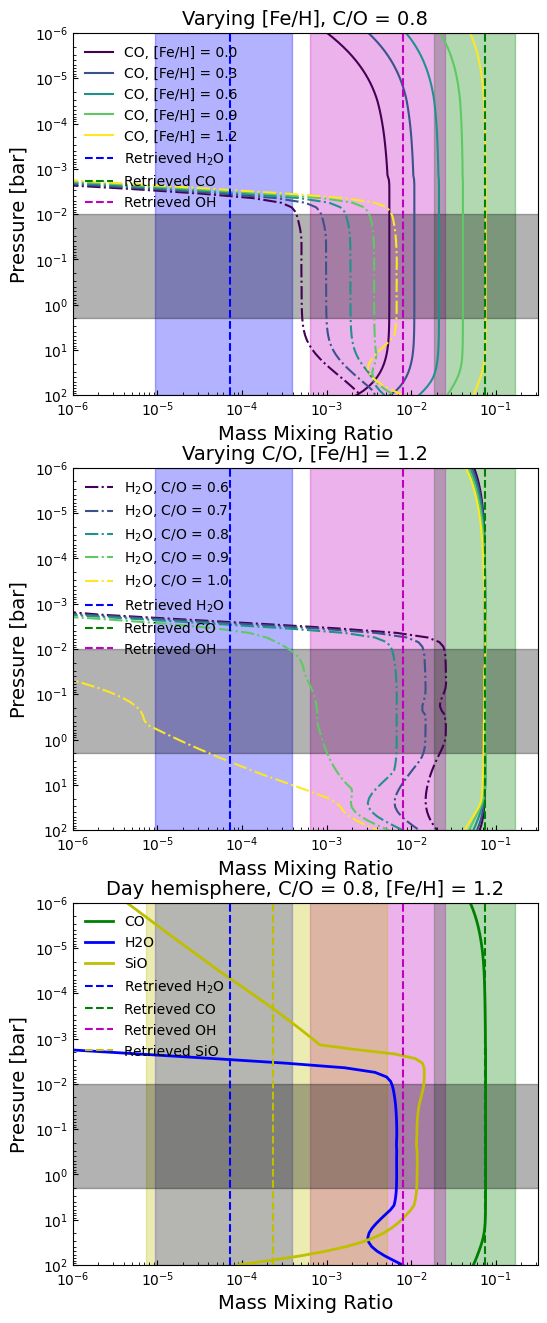}
    \caption{Comparisons between retrieved CO and H$_2$O abundances and abundances interpolated from the \texttt{poor\_mans\_nonequ\_chem} functionality of \petit\ \citep{molliere2017, prt:2019} using the median retrieved PT profile. The shaded pressure layers indicate the region to which our observations are most sensitive based on the emission contribution function in Figure \ref{fig:dayspec}. We compare the retrieved OH abundance to the equilibrium prediction for H$_2$O. These comparisons show the CO abundance is largely insensitive to the C/O ratio, while the H$_2$O abundance is set by both metallicity and C/O ratio. Our results are most consistent with C/O = $0.8\pm0.1$ and $\rm [Fe/H] = 1.2\pm1$, with significant uncertainty in metallicity due to the lack of continuum information. These models predict large SiO abundances, which may be detectable in $L$-band observations. }
    \label{fig:chemplot}
\end{figure}

The retrieved abundances yield a gas-phase $C/O = 0.8^{+0.1}_{-0.2}$. The derived posteriors for C/H, O/H, and C/O are shown in Figure \ref{fig:CO_CH}. However, these retrievals are missing several potential oxygen reservoirs in ultra-hot Jupiter atmospheres, whose presence in significant quantities would lower the atmospheric C/O ratio.  Metal oxides including TiO \citep{cont2021, serindag2021} and AlO \citep{vonessen2019} have previously been detected in WASP-33b, neutral atomic oxygen has been detected in the similar UHJ KELT-9b \citep{borsa2021}, and Si has been detected in low-pressure ($\lesssim1$ mbar) optical observations of WASP-33b \citep{cont2022}, suggesting the possible presence of SiO deeper in the atmosphere. Our limited wavelength coverage precludes detection of any of these species, though a significant reduction in C/O as a result would require either significant neutral oxygen or an atmosphere substantially enriched in refractory elements compared with volatiles.

The atmospheric metallicity of WASP-33b is more difficult to infer from these observations than the C/O ratio. The retrieved abundances suggest a moderately metal-enriched atmosphere, but the lack of continuum information in high-resolution observations leads to strong correlations between species abundances, making absolute abundance estimates difficult. Previously, \citet{cont2022} found a solar-metallicity model best explained the observed Si signal, though they used a widely-spaced model grid. In subsequent analysis \citet{cont2022opt} estimated a metallicity of +1.5 dex based on retrievals of several atomic species from optical data, including Ti, Fe, and V, consistent with the retrieved metallicity from the KPIC observations.

In order to better understand the range of possible C/O ratios and metallicities, we compare our retrieved abundances with grids of equilibrium chemistry models from the \texttt{poor\_mans\_nonequ\_chem} package of \petit\ \citep{molliere2017, prt:2019}. These comparisons are plotted in Figure \ref{fig:chemplot}. In our parameterization scheme, the CO abundance is determined primarily by the atmospheric metallicity, while the H$_2$O/OH abundance is significantly impacted by the C/O ratio. As the equilibrium grid does not include OH, we assume the equilibrium H$_2$O abundance as the prediction for the dayside OH abundance.

Figure \ref{fig:chemplot} shows the retrieved abundances are most consistent with equilibrium models having $\rm [Fe/H] = 1.2$ and C/O = 0.8. These values are consistent with \citet{thorgren2019} modeled upper limits on metallicity for a planet the mass of WASP-33b as well as the retrieved metallicity from \citet{cont2022opt} based on optical observations. However, varying the $\rm [Fe/H]$ in the chemical model suggests that our metallicity uncertainty is approximately 1 dex, and we therefore cannot confidently exclude stellar or even sub-stellar atmospheric metallicities for WASP-33b. C/O appears to be better constrained, with a margin $\pm0.1$ consistent with the posterior shown in Figure \ref{fig:CO_CH}. Note that the strong covariance between species abundances suggests that this C/O ratio will hold even if the metallicity varies significantly (see Figure \ref{fig:CO_CH}).

The chemical equilibrium models suggest significant SiO abundances at the pressures probed by our observations. At lower pressures, the SiO abundance drops due to thermal dissociation, consistent with the optical detection of Si by \citet{cont2022}. While our $K$-band observations are not sensitive to SiO, $L$-band observations of the SiO bands from 4--4.2 $\mu$m could measure the SiO abundance. Assuming SiO is the dominant undetected oxygen reservoir at $\sim1-0.01$ bar pressures, this would significantly improve estimates of the overall oxygen abundance. 

\subsection{CO Isotopologue Ratio}\label{ssec:isotope}

The $^{13}\rm CO/^{12}CO$ posterior peaks at $ \sim 10^{-1.6}$, with a tail towards lower values bringing the median to $10^{-1.7}$. While the posterior is somewhat poorly-constrained, the fact that we see even a weak preference is an argument in favor of high $\rm ^{12}CO$ abundances. In simulated observations, mass-mixing ratios as high as $\rm \log CO = -1.5$ do not always lead to a preference in the isotopologue ratio posterior. The peak of this weak preference is consistent with estimated CO isotopologue ratios from protoplanetary disks \citep{woods2009} and is consistent with the $^{13}\rm CO/^{12}CO \sim 10^{-1} - 10^{-1.6}$ value measured for WASP-77Ab in \citet{Line2021}. Constraints on isotopologue ratios as well as abundances may provide a complementary path to probing formation history and physical conditions in exoplanetary atmospheres \citep{molliere2019, zhang2021}.

\subsection{Implications for Planet Formation}\label{ssec:formation}

The high-C/O, potentially high-metallicity atmosphere suggested by our retrievals is consistent with the \citet{Pelletier2021} observations of the hot Jupiter $\tau$ Boo b. These observations conflict with predictions from pebble accretion models that atmospheric metallicity and C/O ratios should be inversely correlated, as secondary pebble accretion is expected to be dominated by oxygen-rich grains \citep{espinoza2017, madhusudhan2017, cridland2019, khorshid2021}. High C/O, high-metallicity atmospheres require formation and/or substantial accretion in an environment that is enriched in both carbon and solids. Such conditions are consistent with models that incorporate pebble drift \citep{booth2017}, formation near CO or CO$_2$ snow lines \citep{oberg2011}, or with pollution by C-rich grains interior to the H$_2$O snow line \citep{chachan2023}.

As discussed in Section \ref{ssec:ctoo}, one possible explanation for this discrepancy is that our $K$-band high-resolution retrievals are simply missing a significant source of oxygen. Figure \ref{fig:chemplot} suggests relatively high SiO abundances for WASP-33b, which we cannot measure with our $K$-band observations. However, comparison with the equilibrium chemistry models suggests including SiO would not substantially decrease the C/O ratio under equilibrium conditions without substantial enrichment of Si relative to other metals (see Figure \ref{fig:chemplot}, third panel) Alternatively, additional oxygen could be present in the atomic form, as \citep{borsa2021} reported from transit observations of KELT-9b. In the future, Keck/HISPEC will offer simultaneous yJHK coverage at R$\sim10^5$ \citep{hispec} with substantially improved throughput compared with KPIC, enabling characterization of a broader range of chemical species and reducing the likelihood of missing significant oxygen or carbon reservoirs.

Missing oxygen could also be the result of errors in linelists or molecular opacity calculations rather than chemistry. A poorly-matched H$_2$O or OH opacity in the retrieval setup could lead to a bias towards lower abundances. However, both this work and \citet{Pelletier2021} retrieved high-C/O atmospheres using different H$_2$O opacity tables and linelists, indicating that such an error would have to be systematic across multiple independent calculations, while simultaneously not significantly impacting the same calculations for CO. This suggests that the problem would have to be related to a common assumption specific to the H$_2$O high-temperature linelists. Computing high-resolution, high-temperature H$_2$O opacities for the $K$-band alone required nearly one year of CPU time using \texttt{exocross} \citep{exocross2018}, limiting our ability to compute and compare multiple opacity tables.

Alternatively, superstellar metallicity and C/O may arise either from accretion near evaporation fronts close to snow lines \citep{oberg2011} or from carbon-rich grain accretion between the carbon soot line and the H$_2$O snow line \citep{chachan2023}. Pebble accretion studies have found grain accretion within the snow line should boost oxygen abundances rather than carbon \citep[e.g.][]{madhusudhan2017, espinoza2017, cridland2019}, but models incorporating pebble drift can produce giant planet atmospheres with both high metallicity and high C/O. In particular, \citet{booth2017} finds that the C/O $\sim1$ and high carbon abundance we observe in WASP-33b is a specific indication that a planet accreted high-metallicity gas near the CO$_2$ snow line, which is located at $\sim 10$ AU for an A-type primary \citep{oberg2011}. The present observed location and polar orbit of WASP-33b may then be explained through the eccentric Kozai-Lidov Effect \citep{naoz2011} after the protoplanetary disk dissipated. Future incorporation of refractory species abundances, such as Si and Fe, would allow the formation location to be specifically identified, as the refractory-to-volatile ratio is a diagnostic of the formation and accretion history \citep{lothringer2021}. The presence of these species in the gas phase in the atmospheres of UHJs such as WASP-33b presents a unique opportunity to obtain a clear understanding of the evolutionary history of an exoplanet population.

\section{Summary and Conclusions} \label{sec:conc}
We present high-resolution ($R\sim35000$) Keck/KPIC $K$-band observations of WASP-33b covering the dayside of the planet. We successfully retrieve CO, H$_2$O, and OH abundances. We find:

\begin{itemize}
    \item WASP-33b has a dayside thermal inversion. Additional observations covering a larger phase range would offer more precise geographic constraints on the inversion. $K$-band data alone offer only weak constraints on the absolute temperature profile. Larger wavelength coverage may be needed to precisely constrain PT profiles.
    
    \item H$_2$O appears to be mostly dissociated into OH on the dayside. Multi-phase $H$-band observations could constrain the extent of this dissociation by measuring variations in OH and H$_2$O features.
    
    \item WASP-33b has a high C/O ratio, C/O = $0.8^{+0.1}_{-0.2}$. While the metallicity is poorly constrained, there appears to be a weak preference for super-stellar metallicities (roughly 2--12$\times$ WASP-33, but with a $\sim$1 dex uncertainty). Chemical models suggest large SiO abundances which may be detectable in the $L$-band, providing an additional check on the metallicity. 
     
    \item The high-C/O ratio, (possibly) high-metallicity atmosphere of WASP-33b suggests that the atmospheric chemistry of WASP-33b has been significantly influenced by the accretion and migration history of the system. One possible scenario is that WASP-33b accreted most of its envelope from just beyond the CO$_2$ snow line and migrated to its present location after the dissipation of the protoplanetary disk through high-eccentricity migration. Alternatively, WASP-33b could have formed between the carbon sublimation and H$_2$O ice lines if it was substantially contaminated by C-rich grains. There are likely to be other scenarios which could also produce a composition compatible with the observations presented here.
    
\end{itemize}


These observations also demonstrate KPIC's ability to characterize unresolved exoplanet atmospheres. While the reduced throughput of KPIC results in a lower SNR than a slit-based spectrograph, the increased line-spread, blaze, and wavelength stabilities from KPIC's single-mode fiber result in a strong cross-correlation detection suitable for atmospheric retrievals. This detection of WASP-33b demonstrates KPIC's capability to characterize unresolved exoplanets for the first time. Future observations of additional hot and ultra-hot Jupiters, coupled with observations of directly imaged systems, will enable population-level understanding of the thermal and chemical environments of giant exoplanet atmospheres.

\begin{acknowledgments}
We thank the anonymous referee whose detailed and insightful comments improved the quality of this paper. This work used computational and storage services associated with the Hoffman2 Shared Cluster provided by UCLA Institute for Digital Research and Education’s Research Technology Group. L.F. thanks Briley Lewis for her helpful guide to using Hoffman2, and Paul Molli\`ere for his assistance in adding additional opacities to petitRADTRANS. 

L. F. is a member of UAW local 2865. L.F. acknowledges the support of the W.M. Keck Foundation, which also supports development of the KPIC facility data reduction pipeline. The contributed Hoffman2 computing node used for this work was supported by the Heising-Simons Foundation grant \#2020-1821.

Funding for KPIC has been provided by the California Institute of Technology, the Jet Propulsion Laboratory, the Heising-Simons Foundation (grants \#2015-129, \#2017-318 and \#2019-1312), the Simons Foundation (through the Caltech Center for Comparative Planetary Evolution), and NSF under grant AST-1611623.

This research has made use of the NASA Exoplanet Archive, which is operated by the California Institute of Technology, under contract with the National Aeronautics and Space Administration under the Exoplanet Exploration Program.

The data presented herein were obtained at the W. M. Keck Observatory, which is operated as a scientific partnership among the California Institute of Technology, the University of California and the National Aeronautics and Space Administration. The Observatory was made possible by the generous financial support of the W. M. Keck Foundation. The authors wish to recognize and acknowledge the very significant cultural role and reverence that the summit of Maunakea has always had within the indigenous Hawaiian community.  We are most fortunate to have the opportunity to conduct observations from this mountain.

\end{acknowledgments}

%

\vspace{5mm}
\facilities{Keck:II(NIRSPEC/KPIC)}


\software{astropy \citep{astropy:2013, astropy:2018},  
          \dynesty\ \citep{speagle2020},
          \texttt{corner} \citep{corner}, \texttt{exocross} \citep{exocross2018},
          \petit\ \citep{prt:2019, prt:2020}}


\clearpage
\appendix
\section{Corner Plots}\label{app:corner}
Full corner plots are included here for completeness. We discuss the retrievals in Section \ref{sec:res}, including poorly constrained and degenerate parameters.

\begin{figure}
    \centering
    \includegraphics[width=0.9\columnwidth]{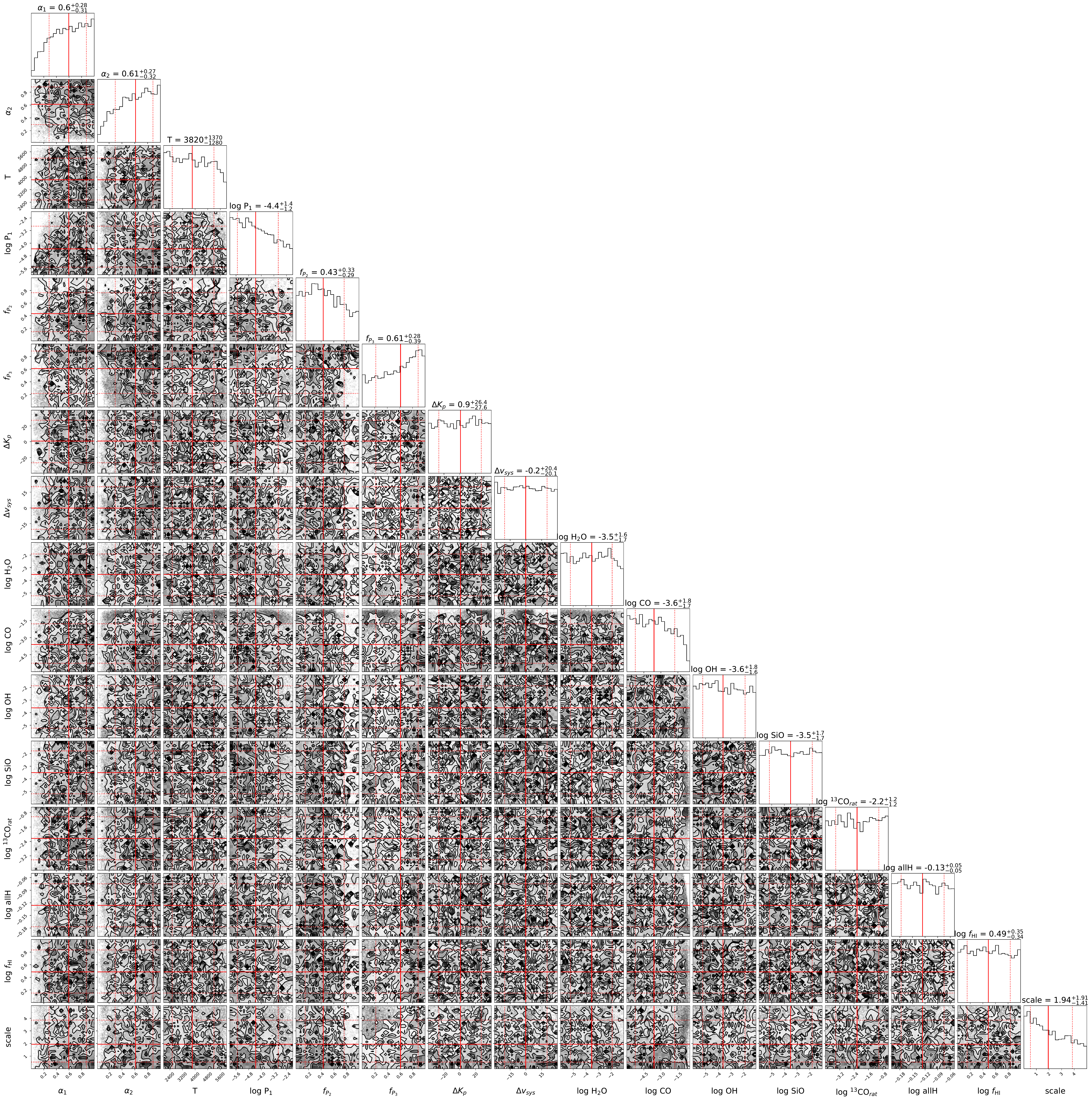}
    \caption{Full corner plot for the test retrieval with no planet. Columns from left to right are the parameters as listed in Table \ref{tab:priors}. Red solid lines indicated the \dynesty\ median, with red dashed lines indicating the bounds of the marginalized 68\% confidence interval. Despite using the same stopping criteria as the other retrievals, none of the posteriors show strong preferences, and the fitter converged in far fewer iterations than when the planet is present in the simulation. The scale parameter shows a weak preference towards smaller values, consistent with the known absence of a planet in the simulated data. The impact of the $T < 6000$ K at all pressure levels requirement can be seen in the behavior of the pressure-temperature profile parameters in the first six columns}
    \label{fig:nocorner}
\end{figure}

\begin{figure}
    \centering
    \includegraphics[width=0.9\columnwidth]{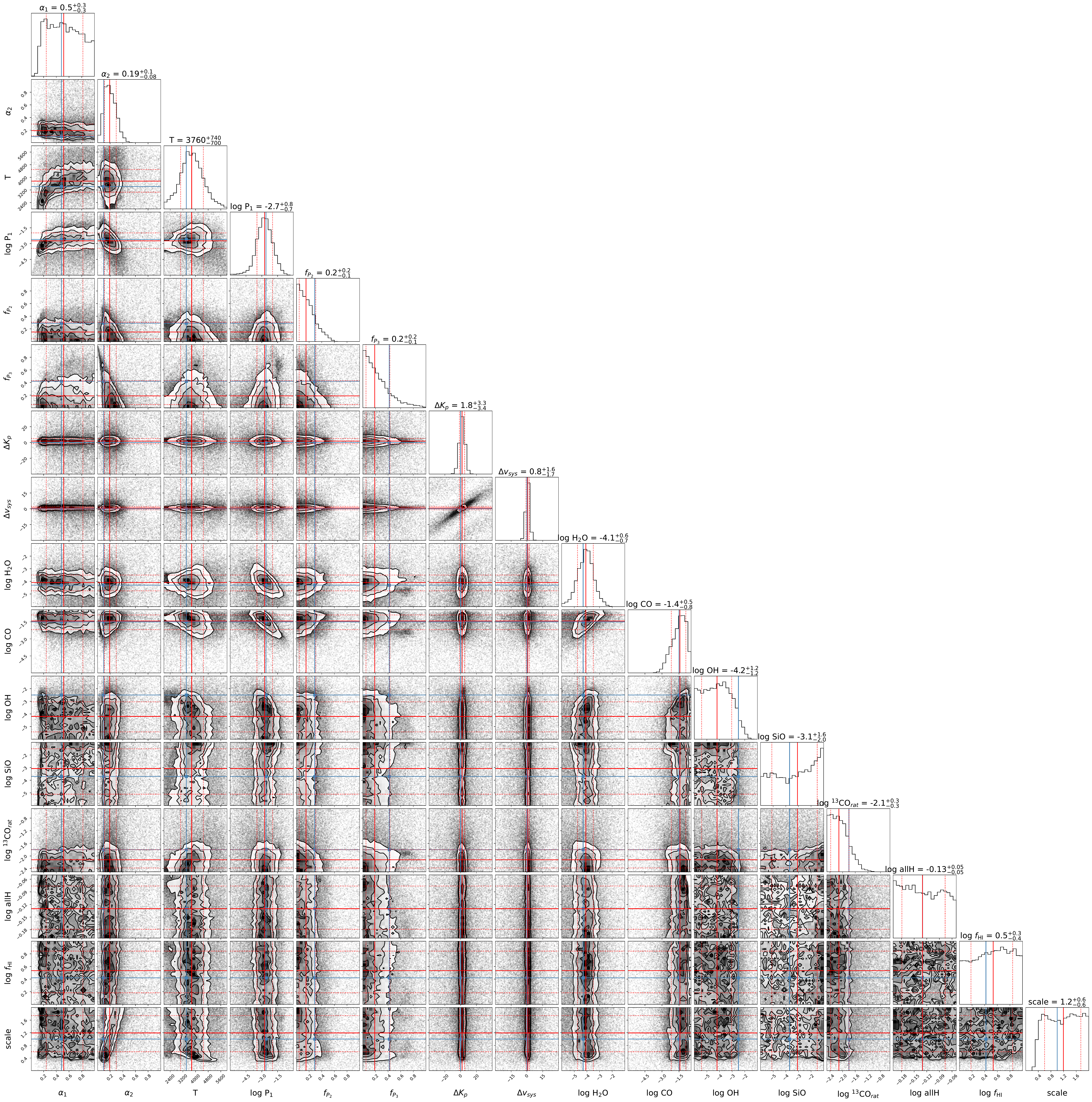}
    \caption{Full corner plot for the test retrieval based on the parameters in Table \ref{tab:simulated}. Columns from left to right are the parameters as listed in Table \ref{tab:priors}. Red solid lines indicated the \dynesty\ median, with red dashed lines indicating the bounds of the marginalized 68\% confidence interval. True values for the simulated data are shown in solid blue. The true values are generally retrieved to within $1\sigma$. }
    \label{fig:corner_test1}
\end{figure}

\begin{figure}
    \centering
    \includegraphics[width=0.9\columnwidth]{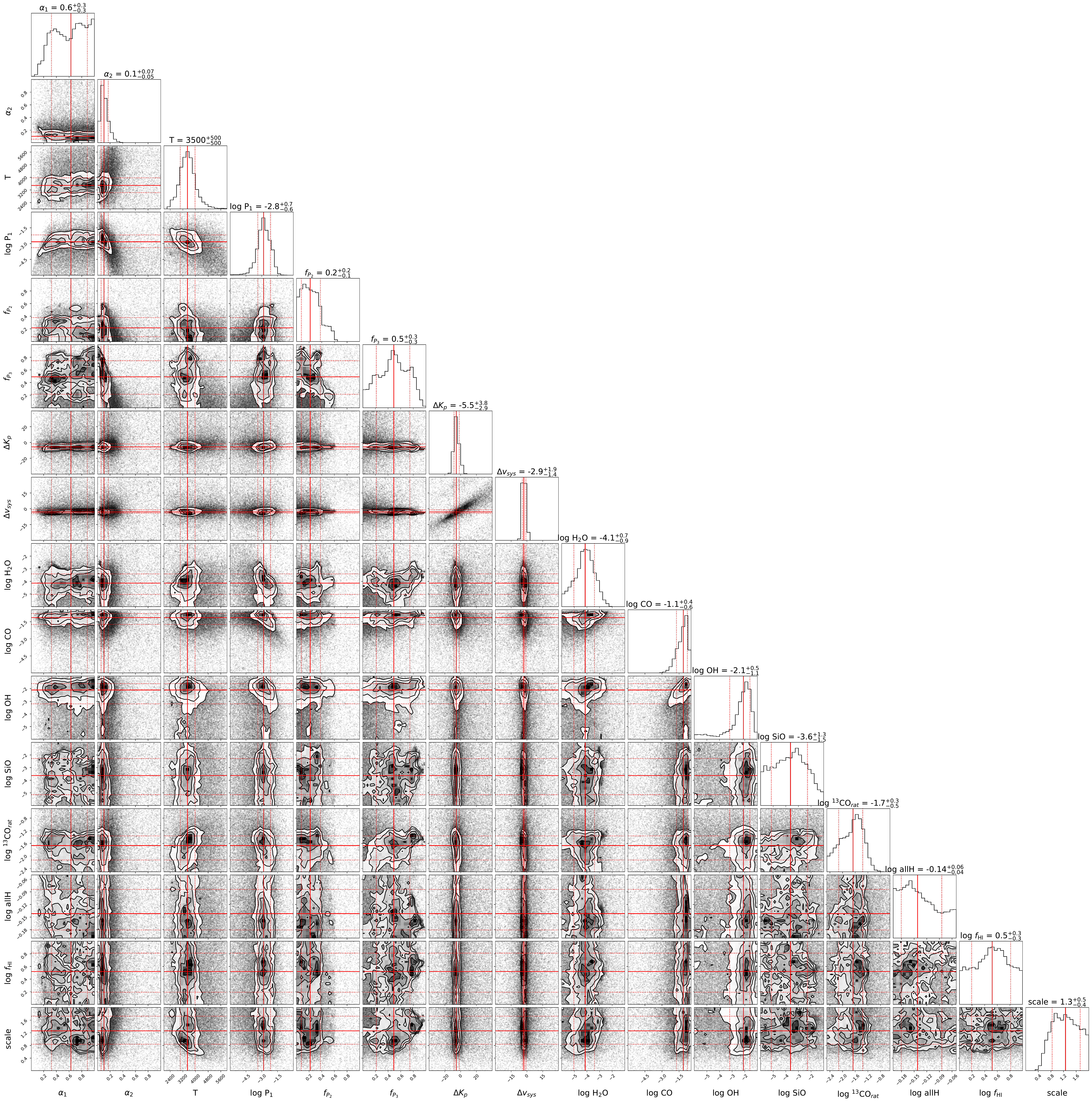}
    \caption{Full corner plot for the WASP-33b retrieval. Columns from left to right are the parameters as listed in Table \ref{tab:priors}. Red solid lines indicated the \dynesty\ median, with red dashed lines indicating the bounds of the marginalized 68\% confidence interval. We discuss these results in Section \ref{sec:res}}
    \label{fig:daycorner}
\end{figure}

\clearpage
\bibliography{exoplanetbib}{}
\bibliographystyle{aasjournal}



\end{CJK*}
\end{document}